\documentclass[aps,pre,twocolumn,groupedaddress,showpacs,nofootinbib]{revtex4-1}
\usepackage[normalem]{ulem}
\usepackage{graphicx}
\usepackage{color}
\usepackage{bbold}
\usepackage{amsmath}
\usepackage{hyperref}
\usepackage[titletoc, title, toc]{appendix}
\usepackage{etoolbox}
\usepackage[space]{grffile}
\usepackage[caption=false]{subfig}

\newcommand{\nn}{\nonumber}
\newcommand{\hs}{\hspace{0.25cm}}
\newcommand{\svd}{\vdots \hspace{0.05cm}}

\DeclareMathOperator{\T}{T}
\DeclareMathOperator{\Tr}{Tr}

\newcommand{\der}{\partial}
\newcommand{\id}{\mathbb{1}}
\newcommand{\C}{\mathcal{C}}
\newcommand{\J}{\mathcal{J}}
\newcommand{\ab}{_{\alpha \beta}}
\newcommand{\F}{\mathcal{F}}
\newcommand{\Fab}{\mathcal{F}\ab}
\newcommand{\G}{\mathcal{G}}
\newcommand{\rv}{\vec{r}}
\newcommand{\fv}{\vec{f}}
\newcommand{\gv}{\vec{g}}
\newcommand{\Rv}{\vec{R}}
\newcommand{\rhov}{\vec{\rho}}
\newcommand{\rhotv}{\vec{\tilde{\rho}}}
\newcommand{\Cexp}{\langle \C \rangle}
\newcommand{\Vexp}{\langle V \rangle}
\newcommand{\sexp}{\langle \sigma_{nn} \rangle}
\newcommand{\gexp}{\langle \gv \cdot \gv \rangle}
\newcommand{\gxexp}{\langle g_x \rangle}
\newcommand{\gixexp}{\langle g^i_x \rangle}

\newcommand{\Nst}{N_{\rm ST}}

\newcommand{\Ttotal}{{\bf \Theta}}
\newcommand{\lavg}{\langle l \rangle}

\begin{document}

\title{Affine and topological structural entropies in granular statistical mechanics: \\
explicit calculations and equation of state}
\author{Shahar Amitai$^1$ and Raphael Blumenfeld$^{1,2,3}$}

\affiliation{1. Imperial College London, London SW7 2BP, UK \\
2. College of Science, NUDT, Changsha, 410073 Hunan, PRC \\
3. Cavendish Laboratory, Cambridge CB3 0HE, UK}

\date{\today}

\begin{abstract}

We identify two orthogonal sources of structural entropy in rattler-free granular systems - affine, involving structural changes that only deform the contact network, and topological, corresponding to different topologies of the contact network. We show that a recently developed connectivity-based granular statistical mechanics separates the two naturally by identifying the structural degrees of freedom with spanning trees on the graph of the contact network. We extend the connectivity-based formalism to include constraints on, and correlations between, degrees of freedom as interactions between branches of the spanning tree. We then use the statistical mechanics formalism to calculate the partition function generally and the different entropies in the high-angoricity limit. We also calculate the degeneracy of the affine entropy and a number of expectation values. From the latter, we derive an equipartition principle and an equation of state relating the macroscopic volume and boundary stress to the analogue of the temperature, the contactivity.

\end{abstract}


\maketitle

\section{Introduction}

Granular matter is one of the most significant forms of matter in nature, both on Earth and celestially. It is also relevant to human society in many ways, be it in the context of products and technological applications or through our interactions with the natural environment around us. Yet, a reliable fundamental understanding of this form of matter is yet to emerge, a situation that is limiting the development of predictive and effective modelling. Consequently, this area has been the focus of intensive research in recent years.

One of the main modelling tools in the theorist's arsenal is statistical mechanics. This powerful method, devised originally to deal with thermodynamic systems subject to thermal fluctuations, is the ultimate coarse-graining technique. Statistical studies of granular assemblies date back to the 1920s \cite{Sm29}, but the introduction of granular statistical mechanics (GSM) in 1989 \cite{EdOa89, *EdOa89_2, MeEd89} led to a significant increase in research in this direction within the physics community. A number of major advances include the calculation of the volume function \cite{BlEd03}, the introduction of the stress ensemble \cite{EdBl05, Heetal07}, the finding that the two ensembles are interdependent \cite{Bletal12} and the measurement of either ensemble's equilibration \cite{PuDa13}. The potential advantage of statistical mechanics is in the ability to derive with it equations of state and constitutive relations, a holy grail in the field. Yet, such relations have been slow to emerge for a variety of reasons \cite{Bl07, BlEd09, Bletal15}.

GSM is based on entropy, namely the number of structural and stress configurations that static assemblies of macroscopic particles can have \cite{Bletal05, BlEd06, Puetal10}. Recently, the entropy of packs of up to $N = 128$ soft particles was measured numerically and found to be extensive after the subtraction of $\ln N!$. \cite{Asetal14, Maetal16, Maetal16_2}

\begin{figure}[h]
\includegraphics[width=0.45\textwidth]{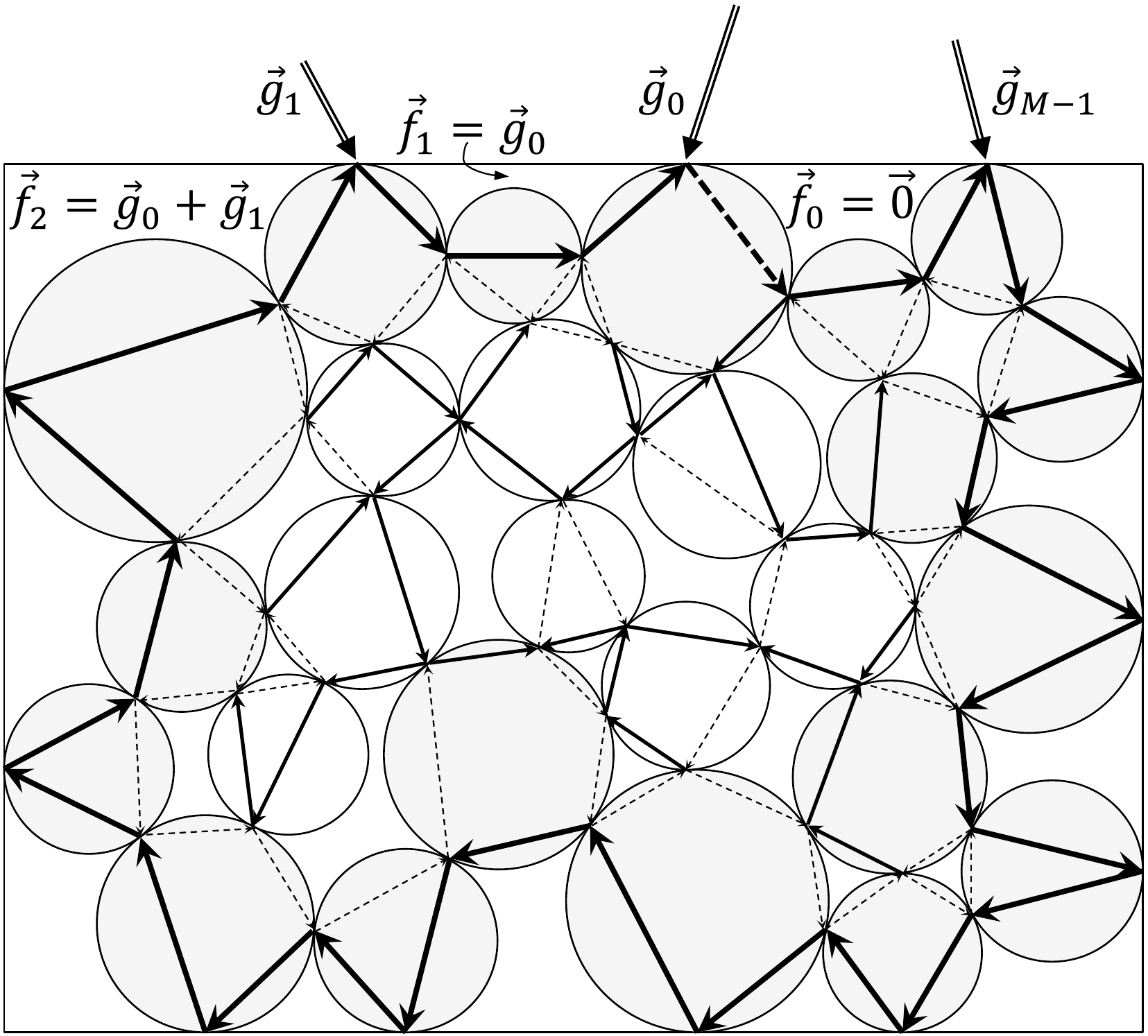}
\caption{A 2D poly-disperse granular pack with $(\alpha - 1) M = 18$ boundary particles (slightly shaded), of which $M = 12$ touch the walls (and thus $\alpha = 2.5$). The vectors $\rv$ (solid and dashed, thick and thin) connect a particle's nearest contacts, circulating clockwise. Their components constitute the vector $\Rv$. The solid vectors (both thick and thin) form a non-directional spanning tree: they form no loops, representing the independent DFs, and they reach every contact point. Their components constitute the vector $\rhov$. Each dashed vector is a linear combination of solid vectors, namely one can always get from its tail to its head by walking along solid vectors. There are $\alpha M = 30$ boundary (thick) vectors $\rv$, and our choice of spanning tree includes all of them but one. Overall the pack has $N_c = 56$ contacts, and thus $N_\rho = 55$ independent (solid) vectors $\rv$. Also illustrated are three boundary forces, $\gv^b$, and three boundary loop forces, $\fv^b$. As explained in the text, we set $\fv^0 = 0$, and the rest are accumulation of the boundary forces, going anti-clockwise around the pack.}
\label{fig:pack}
\end{figure}

In general, GSM consists of two sub-ensembles, one of all the structural configurations and the other of all stress microstates. Originally, the structural microstates were proposed to occur with probability that depends on their volume \cite{EdOa89, *EdOa89_2, MeEd89}, but it was shown recently that this formulation is flawed \cite{Bletal16}, which may have also been responsible for the little use of the GSM in the community. A new formulation, based on a connectivity function, was then proposed, but it has not been tested yet. The connectivity-based partition function is
\begin{align} \label{eq:z_orig}
Z = \int &\prod_{n = 1}^{N_c - 1} d \rv_n \prod_{m = 1}^{M} d \gv_{m} \cdot \nn \\
&\Ttotal ( \{ \rv \} ) \cdot G ( \{ \rv \} ) \cdot e^{- \frac{\C \left( \{ \rv \} \right)}{\tau} - \gamma : \F \left( \{ \rv \}, \{ \gv \} \right)} \ ,
\end{align}
where the vectors $\{\rv\}$ connect contacts around particles and $\{\gv\}$ are the compressive forces acting on the boundary particles (see figure \ref{fig:pack} for an example in two dimensions (2D)). $\tau = \partial \Cexp / \partial S$, with $S$ the entropy, is the `contactivity' - a measure of the connectivity fluctuations that is an analogue of the temperature. $\F$ is the force moment tensor, formed by the outer product of the intergranular forces and their position vectors, summed over all contacts. This function couples the structure and the stress ensembles \cite{EdBl05, Heetal07, Bletal12}. $\gamma\ab = 1 / X\ab$ is an (inverse) angoricity tensor \cite{EdBl05} and $\Ttotal$ includes the constraints on the systems forming the ensemble, e.g. that they are in mechanical equilibrium, generated by the same process and all have the same mean coordination number $\bar{z}$. $G$ is called the measure, which is independent of the exponential Boltzmann factor and represents the probability to sample a specific configuration of $\rv$-vectors. The measure may depend on the preparation and sampling protocols, and need not be uniform, as recent results suggest \cite{Asetal14, Maetal16, Maetal16_2}. Its form is not known for our systems and, for simplicity, we take it to be uniform. Nevertheless, the following analysis can be carried out for any form of $G$. The connectivity, $\C = \sum_q \rv^q \cdot \rv^q$, is a sum over all $N_r$ inter-contact vectors. Of these vectors, only $N_c - 1$ are independent, where $N_c$ is the number of contacts. In principle, the structural degrees of freedom (DFs) should also include the parameters specifying the particle shapes. These are ignored here for brevity, but could be included without loss of generality \cite{Bletal15}.

In this paper, we demonstrate the use of the full structure-stress partition function, whose structural part is based on the recent connectivity function. The paper is constructed as follows: In section \ref{sec:partition_function} we rewrite the partition function of eq. (\ref{eq:z_orig}) in a convenient form. In section \ref{sec:entropy} we clarify the distinction between two different types of entropies in granular materials: affine and topological. In section \ref{sec:calculations} we calculate expectation values, and the affine entropy. We also derive an equipartition principle and an equation of state. In section \ref{sec:interactions} we show that the connectivity partition function readily accommodates structural constraints and correlations between the vectors $\rv$. In section \ref{sec:topological} we outline the calculation of the topological entropy. We conclude in section \ref{sec:discussion}.

\section{The partition function} \label{sec:partition_function}

The vectors $\rv$ are the structural DFs. In 2D, they run clockwise around each particle (see figure \ref{fig:pack}) and there are $N \bar{z}$ of them. A similar parameterisation exists in three dimensions (3D) \cite{BlEd06, Fretal08}. For clarity, we focus here on 2D systems experiencing no body forces. We denote by $M$ the number of boundary particles to which external compressive forces are applied, and by $(\alpha - 1) M$ ($\alpha = O(1)$) the total number of boundary particles. It follows that $\alpha M$ is the number of the outermost boundary vectors $\rv$ in the system.

In 2D, the vectors $\rv$ form loops around particles and around cells and, therefore, only $N_\rho = N_c - 1$ of them are independent. The set of independent vectors $\rv$, of which there are many possibilities, forms an undirected spanning tree on the contact network \cite{Jo14}. As we will see below, it is convenient to constrain our choice of spanning tree to include $\alpha M - 1$ of the outermost boundary vectors and denote those by $\rv^b$. For brevity, we define $\Rv$ as the ($d \times N_r$ long) vector, containing the components of all the vectors $\rv$. The entries of $\Rv$ are ordered as follows: First come the $\alpha M - 1$ independent boundary vectors $\rv^b$ ($d$ entries each), then $N_c - \alpha M$ independent vectors $\rv$ in the bulk, and then the remaining (dependent) vectors $\rv$. We can then write
\begin{align}
\Rv = A \rhov \ ,
\end{align}
where $\rhov$ is a $d \times N_\rho$ long vector, containing only the components of the independent vectors $\rv$. The top part of $A$ is clearly a unit matrix, whereas each line of the bottom part describes the `route' along the spanning tree to get from the tail to the head of a certain dependent $\rv$. It follows that all the entries of $A$ are predominantly $0$, with fewer $\pm 1$ entries. In terms of $\rhov$ we have
\begin{align} \label{eq:C}
\C / \tau = \rhov A^T A \rhov / \tau = \frac{1}{2} \rhov B \rhov \hs \hs (B \equiv \frac{2}{\tau} A^T A) \ .
\end{align}

The stress ensemble is controlled by the force moment tensor, which can be written either as a sum over the outer product of the contact force and position at every contact point, or in terms of the loop forces, $\fv^i$, defined in \cite{BaBl02}, namely, $\Fab = \sum f^i_\alpha r^i_\beta$. While the sum is over all the cells, the contribution of internal cells vanishes. This is because $\fv^i$ is constant for every cell, and the sum over the vectors $\rv$ that enclose the cell vanishes. It follows that the only contribution to $\F$ comes from the boundary vectors, $\sum_{b=1}^{\alpha M} f^b_\alpha r^b_\beta$. These vectors are all, but one, independent and constitute the first entries of $\rhov$. The loop forces $\fv^b$ are accumulation of the boundary forces $\gv^b$ (see figure \ref{fig:pack}). Recalling that the loop forces are only defined up to a constant \cite{BaBl02}, we set the loop force, associated with the single dependent boundary vector, to zero. We can, thus, manipulate the force moment term in the partition function to the form
\begin{align} \label{eq:Fab}
\gamma : \F = \gv H^T \gamma E \rhov = \gv Q^T \rhov \hs (Q \equiv E^T \gamma^T H) \ .
\end{align}
with $E$ and $H$ defined in eq. (\ref{eq:CandE}) below. Noting that the matrices $A$, $B$, $E$ and $H$ do not involve the Cartesian coordinates of the matrix $\gamma$, the latter could be placed as shown in (\ref{eq:Fab}). This position is arbitrary with respect to $H$ or $E^T$ in the definition of $Q$.

\begin{widetext}
\begin{align} \label{eq:CandE}
& \hspace{0.7cm} {\rm boundary} \hspace{0.35cm} {\rm boundary} \hspace{0.45cm} {\rm boundary} \hspace{1.5cm} {\rm boundary} \hspace{0.55cm} {\rm bulk} \nn \\
& \hspace{1cm} {\rm cell \ 1} \hspace{1.05cm} {\rm cell \ 2} \hspace{1.05cm} {\rm cell \ 3} \hspace{1.9cm} {\rm cell} \ M \! - \! 1 \hspace{0.4cm} {\rm vectors} \nn \\
H = \left( \begin{array}{cccccccc}
0 & 0 & 0 & 0 & 0 & ... & 0 & 0 \\
1 & 0 & 0 & 0 & 0 & ... & 0 & 0 \\
1 & 1 & 0 & 0 & 0 & ... & 0 & 0 \\
1 & 1 & 1 & 0 & 0 & ... & 0 & 0 \\
\vdots & \vdots & \vdots & \vdots & \vdots & \ddots & \vdots & \vdots \\
1 & 1 & 1 & 1 & 1 & ... & 1 & 0 \end{array} \right) \hs ; \hs
E = &\left( \begin{array}{rrr|rr|rrr|r|rr|r}
\hs 0 & \hs 0 & \hs 0 \hs &    -1 &     -1 \hs &    -1 &     -1 &     -1 \hs &                    &     -1 &    -1 \hs & \\
      1 &       1 &       1 \hs &     0 &      0 \hs &     0 &      0 &      0 \hs &                     &      0 &     0 \hs & \\
      0 &       0 &       0 \hs &     1 &      1 \hs &     0 &      0 &      0 \hs &                     &      0 &     0 \hs & \\
      0 &       0 &       0 \hs &     0 &      0 \hs &     1 &      1 &      1 \hs & \hs \dots \hs &      0 &     0 \hs & \hs \hs 0 \hs \hs \\
 \svd &   \svd &  \svd \hs & \svd & \svd \hs & \svd & \svd & \svd \hs &                      & \svd & \svd \hs & \\
      0 &       0 &       0 \hs &     0 &      0 \hs &     0 &      0 &      0 \hs &                      &      1 &     1 \hs & \end{array} \right)
\end{align}
\small The matrices $H$ and $E$, used in eq. (\ref{eq:Fab}), and corresponding to the system in figure \ref{fig:pack}. $H$ is accumulation matrix, transforming from the boundary forces, $\gv$, into the loop forces, $\fv = H \gv$, where we set $\fv^0 = 0$. $E$ transforms from the independent vectors, $\rhov$, into the boundary vectors, $\rv^b$, and sums the boundary vectors $\rv^b$ of each boundary cell. These are always several consecutive vectors from the $M-1$ first entries of $\rhov$. Since the first cell includes one dependent boundary vector, the sum of this cell's vectors is the (negative) sum of all the other boundary vectors $\rv^b$.
\end{widetext}

\section{Affine and topological structural entropies} \label{sec:entropy}

Before we continue, it is essential to identify and classify three different sources contributing to the entropy of granular systems,
\begin{align} \label{eq:AllEntropy}
S = S_r + S_a + S_t \ .
\end{align}
(i) $S_r$ is the {\it rattler entropy}, which consists of all the positional variations of the rattlers inside voids. The rattlers are particles that, in the absence of external and body forces, are not part of the force-carrying structure. This contribution is essentially the volume that these rattlers can occupy within any given structure and as such it is straightforward to calculate. Nevertheless, it has little bearing on most of the mechanisms governing the physics of static granular assemblies, which involve force transmission between particles. Therefore, this contribution is ignored in the following discussion. \\
(ii) $S_a$ is the {\it affine entropy}. Regarding the contact network as vertices of a graph, whose edges are the vectors $\rv$, $S_a$ consists of all the possible affine, i.e. connectivity-preserving, distortions of this graph, which the given collection of $N$ force-carrying particles can make. We shall include the entropy coming from the stress DFs in $S_a$ as well. \\
(iii) $S_t$ is the {\it topological entropy}, which consists of all the possible different topologies, or connectivity networks, that the same collection of particles can make.

Eq. (\ref{eq:AllEntropy}) is based on the assumption that the three entropies are additive, i.e. that the partition function is a product of three independent parts, $Z = Z_{\rm r} \cdot Z_a \cdot Z_{\rm t}$. This mean field-like approximation is justified in section \ref{sec:topological}. Note that the distinction between $S_a$ and $S_t$ is useful mainly when the mean coordination number is kept fixed across the systems in the ensemble, which is implicitly assumed here. Letting it fluctuate gives rise to different topologies, obviating the affine entropy. As we will show below, the distinction between these two types of entropy is crucial because their scaling with $N$ is markedly different.

\section{Affine entropy - calculation, expectation values and equation of state} \label{sec:calculations}

All the connectivity, or topological, information is contained in the matrix $A$ in the sense that the connectivity of a specific configuration corresponds to a specific set of entries $A_{ij}$. Thus, $S_a$ is the result of the different changes in the vectors $\rv$, under the constraint that the matrix $A$, and hence $B$, are fixed. To calculate the affine entropy, we use eq. (\ref{eq:C}) and (\ref{eq:Fab}) to express the partition function of eq. (\ref{eq:z_orig}) in terms of $\rhov$ and $\gv$:
\begin{align} \label{eq:z_simple}
Z_a = \int e^{- \frac{1}{2} \rhov B \rhov - \gv Q^T \rhov} (d \rhov)(d \gv) \ .
\end{align}

In the following, we first calculate $Z_a$ while ignoring the constraints and taking $\Ttotal = 1$. We will then point out the problems that this simplification leads to and adjust the calculation to model in the constraints. The integral in (\ref{eq:z_simple}) is Gaussian and its calculation straightforward. Changing variables, $\rhotv \equiv \rhov + B^{-1} Q \gv$, we have
\begin{align} \label{eq:to_rho_tilde}
- \frac{1}{2} \rhov B \rhov - \gv Q^T \rhov = - \frac{1}{2} \rhotv B \rhotv + \frac{1}{2} \gv P \gv \ ,
\end{align}
with $P \equiv Q^T B^{-1} Q$. The partition function then reduces to two decoupled integrals,
\begin{align} \label{eq:z_square}
Z_a = \int e^{- \frac{1}{2} \rhotv B \rhotv} (d \rhotv) \int e^{+ \frac{1}{2} \gv P \gv} (d \gv) \ ,
\end{align}
which can be calculated readily because $B$ and $P$ are symmetric and hence diagonalisable by orthogonal matrices. $B$ is positive definite, since for any choice of $\rhov \ne \underline{0}$ we have $\rhov B \rhov = \Rv \cdot \Rv > 0$, which means that all its eigenvalues are positive. The matrix $P$, however, is singular -- it has $d$ zero eigenvalues, leading to an integrand of 1. The other eigenvalues are positive, leading to an {\it increasing exponential} due to the positive sign in eq. (\ref{eq:z_square}). This, however, does not lead to a diverging integral because the boundary forces, $\gv$, are finite. The integrals in (\ref{eq:z_square}) are straightforward to calculate,
\begin{align} \label{eq:z_final}
Z_a = \sqrt{\frac{(2 \pi)^{d N_\rho} 2^{3 d (M - 1)}}{|B|^d |P|_+}} (2 g)^d e^{\frac{g^2}{2} \Tr(P)} \prod_{p_i > 0} D(a_i) \ ,
\end{align}
where $|B|$ is the determinant of $B$, $p_i$ are the eigenvalues of $P$, $|P|_+$ is its pseudo-determinant (i.e. the product of all non-zero $p_i$'s), $g$ is the maximal boundary force, $a_i \equiv g \sqrt{p_i / 2}$ and $D$ is the Dawson function:
\begin{align} \label{eq:dawson}
D(\alpha) \equiv e^{-\alpha^2} \int_0^\alpha e^{x^2} dx \ .
\end{align}

We can now use eq. (\ref{eq:z_final}) to calculate the affine entropy and several expectation values, for a given contact network. The detailed calculations are shown in appendix \ref{appendix:calc} and here we present the main results. \\

\noindent 1. The mean connectivity can be calculated using $\Cexp_a = \tau^2 (\der \ln Z_a / \der \tau)$, while keeping $A$ constant,
\begin{align} \label{eq:cexp}
\Cexp_a = \frac{\tau}{2} \left[ d (N_\rho - M + 1) + \sum_{a_i > 0} \frac{a_i}{D(a_i)} \right] \ .
\end{align}
In the high angoricity limit, $a_i / D(a_i) = 1$, and since there are $d(M-1)$ of them, eq. (\ref{eq:cexp}) becomes
\begin{align} \label{eq:EqPart}
\Cexp_a = N_\rho d \tau / 2 \ .
\end{align}
This is an equipartition principle \cite{Bletal16} -- the mean connectivity is shared among the $d N_\rho$ structural DFs, with each getting on average $\tau/2$, analogously to the mean energy of $k_B T/2$ per DF in thermal systems. Eq. (\ref{eq:cexp}) deviates from this by $O(M \sim \sqrt{N} \ll N)$, since the sum on the right is over $d(M-1)$ elements of $O(1)$.

In the low angoricity limit, $D(a_i) \approx 1/2a_i$, and the sum on the right hand side of (\ref{eq:cexp}) becomes $g^2 Tr(P)$. This is a sum over $O(M)$ finite terms and it scales as $\tau \Tr (\gamma^2)$. Thus, it is also negligible relative to the first term as long as $g^2 \tau \Tr(\gamma^2) < O(M)$, and equipartition holds in this regime too up to terms of $O(\sqrt{N})$. \\

\noindent 2. The mean squared-norm of the force vector is
\begin{align} \label{eq:gexp}
\gexp_a = -\Tr(\Pi) + \sum_{a_i > 0} \frac{g^2}{2 a_i D(a_i)} \ ,
\end{align}
where $\Pi$ is the pseudo-inverse matrix of $P$, $P \Pi P = P$. Note that $P \Pi \ne I$, as $P$ is singular. The statistics of the boundary forces should be hardly dependent on the precise internal topology and, hence, $\gexp_a = \gexp$. The first term on the right hand side scales as $1 / [\tau \Tr (\gamma^2)]$. The dependence of the second term on $\tau$ and $\gamma$ is more complex as it is via the non-linear Dawson function. However, this dependence can be obtained in two limits. In the high angoricity limit $D(a_i) \approx a_i$ and the dependence is the same, $1 / [\tau \Tr (\gamma^2)]$. In the low angoricity limit, $D(a_i) \approx 1/2a_i$, the second term approaches a constant and its dependence on $\tau$ and $\gamma$ disappears. \\

\noindent 3. Taking the system to be a square of $\sqrt{N} a \times \sqrt{N} a$, where $a$ is the typical size of a particle, and assuming isotropic boundary stresses, the total force normal to one side is $M \gxexp_a / 4$, with $g_x$ the component of a boundary force normal to the wall. An explicit calculation of the expectation value $\gxexp_a$ (see appendix \ref{appendix:calc}) gives $\gxexp_a = |\G| / \sqrt{M}$, with $\G$ an $M$-long vector, whose components are $\G_i = g (1 - e^{- a_i^2}) / (2 a_i D(a_i))$. Summing over all the normal components of the external forces along the side and dividing by the length, we obtain the expectation value of the normal stress
\begin{align} \label{eq:sexp}
\sexp_a = \frac{M}{4 \sqrt{N} a} \cdot \frac{ |\G| }{\sqrt{M}} \approx \frac{ |\G| }{4 a \sqrt[4]{N}} \ .
\end{align}

The components of $\G$ range between 0.5 and 1, as we show explicitly in appendix \ref{appendix:order_of_curly_d}, leading to $|\G| = O(\sqrt{M} \approx \sqrt[4]{N})$. Therefore, $\sexp_a$ is independent of system size, as expected. \\

\noindent 4. The affine entropy, given by $S_a = \Cexp_a / \tau + \ln Z_a$, is
\begin{align} \label{eq:entropy}
S_a = &- \frac{d}{2} \ln |B| - \frac{1}{2} \ln |P|_+ + \frac{g^2}{2} \Tr(P) \nn \\
&+ \sum_{a_i > 0} \left[\ln D(a_i) + \frac{a_i}{2 D(a_i)} \right] \\
&+ \frac{d}{2} \left[ N_\rho \ln (2 \pi e) + 2 \ln (2 g) + (M - 1) \ln \left( \frac{8}{e} \right) \right] \ . \nn
\end{align}
There is more to this result than being satisfyingly exact and testable - it also provides the following significant observation. The terms involving $|B|$ and $N_\rho$ are of order $N$, thus dominating over all the other terms, which are of order $M\sim\sqrt{N}$. Since these two terms originate only in the structure they are independent of the external boundary forces. This means that, for sufficiently large systems, {\it the boundary forces contribute negligibly to the affine entropy regardless of the angoricity value}. Consequently, equation (\ref{eq:entropy}) substantiates the generality of previous results obtained in the high angoricity limit. Keeping only these two terms, we obtain
\begin{align} \label{eq:AffEntropy}
S_a \approx \frac{d}{2} \left[ N_\rho \ln (2 \pi e) - \ln |B| \right] \ .
\end{align}
This result was also derived and calculated numerically in \cite{Bletal16}. In particular, it was shown there  to scale linearly with $N$. This makes the affine entropy conveniently extensive. It also increases with $\tau$, the measure of structural fluctuations, as expected. In appendix \ref{appendix:approx} we discuss in more detail different approximations for $S_a$. \\

\noindent 5. To calculate the mean volume, note that the volume is a quadratic function of all the independent vectors $\rv$, $V \equiv \rhov W \rhov$. Rewriting it in terms of the transformed variables of eq. (\ref{eq:to_rho_tilde}), we have
\begin{align} \label{eq:vexp_orig}
V &= \rhov W \rhov = \rhotv W \rhotv - 2 \gv Q^T B^{-1} W \rhotv + g U g \ ,
\end{align}
with $U \equiv Q^T B^{-1} W B^{-1} Q$. The cross term vanishes on integration over the symmetric distribution of $\rhov$, giving
\begin{align}
\Vexp_a &= \frac{1}{Z_a} \int (\rhotv W \rhotv + g U g) e^{- \frac{1}{2} \rhotv B \rhotv + \frac{1}{2} \gv P \gv} (d \rhotv)(d \gv) \nn \\
&= d \Tr(W B^{-1}) - \Tr(U \Pi) + \sum_{a_i > 0} \frac{g^2 U_{ii}'}{2 a_i D(a_i)} \ , \label{eq:vexp}
\end{align}
with $U'$ the matrix $U$, rotated to the basis where $P$ is diagonal (See appendix \ref{appendix:calc} for details). 
The first two terms in eq. (\ref{eq:vexp}) are linear in $\tau$ and are independent of $\gamma$, but the third term depends on both $\tau$ and $\gamma$. As in eq. (\ref{eq:gexp}), this dependence is not simple. However, in the high angoricity limit it becomes also linear in $\tau$, and at low angoricity it scales as $\tau^2 \Tr (\gamma^2)$. The expectation value $\Vexp_a$ suffers from a problem, which we discuss in detail and then resolve in the next section. \\

\noindent 6. To obtain the affine contribution to the equation of state, we combine eqs. (\ref{eq:sexp}) and (\ref{eq:vexp}) to obtain
\begin{align} \label{eq:EOS}
\sexp_a \Vexp_a = F(\tau, X\ab, g) \ .
\end{align}
The form of this expression resembles that of the ideal gas equation of state, $P V = N k_B T$: the left hand side involves two macroscopically measurable quantities, while the right hand side is a function of the `temperature-like' variables alone, the angoricity and contactivity.

The above expectation values and equation of state, (\ref{eq:cexp}), (\ref{eq:gexp}), (\ref{eq:sexp}), (\ref{eq:entropy}), (\ref{eq:vexp}) and (\ref{eq:EOS}), can be simplified by approximating the Dawson function for low and high angoricities, as we show in detail in appendix \ref{appendix:approx}. In particular, in the high angoricity limit, we obtain the following explicit equation of state,
\begin{align}
\sexp_a \Vexp_a \approx \frac{g}{8} M \left[ d \Tr(W B^{-1}) + \frac{g^2}{3} \Tr(U \Pi P) \right] \ .
\end{align}
In this expression, the first term on the right hand side scales as $\tau$ and the second term scales as $\tau^2 \Tr(\gamma^2)$. In this limit, the second term, which originates in the forces entropy, is negligibly small compared to the first, in agreement with our observation, eq. (\ref{eq:entropy}).

\section{Extending the formalism: including constraints and vector-vector interactions} \label{sec:interactions}

A close scrutiny of the result for $\Vexp_a$, eq. (\ref{eq:vexp}), reveals a problem - the expected volume vanishes. To see this, we introduce a projection operator onto the Cartesian coordinates, $\J$. Using this operator on $B$, which does not depend on the Cartesian coordinates, gives $\J(B) = \id$, with $\id$ the 2D identity matrix; $\J(W) = \epsilon$, with $\epsilon$ the 2D Levi-Civita symbol, since volume is a sum of cross products; $\J(P) = \gamma \cdot \gamma^T$; and $\J(U) = \gamma \cdot \epsilon \cdot \gamma^T$. Using these, the first term on the right hand side of eq. (\ref{eq:vexp}) vanishes, $\Tr(W B^{-1}) \propto \Tr(\J(W B^{-1})) = \Tr(\epsilon) = 0$. The second term also vanishes because $\Tr(U \Pi) \sim \Tr(\epsilon) = 0$. Similar considerations give that $U_{ii}' = 0$, which means that the third term also vanishes and we obtain that overall $\Vexp_a = 0$.

This, of course, cannot be correct. The problem can be traced to the omission of the function $\Ttotal$ from the partition function, which allowed the independent vectors, $\rv$, to be unconstrained in the integral (\ref{eq:vexp}). In particular, for every occurrence of $\rv$ there is an occurrence of $-\rv$. This means that, for every structural configuration we consider, there is a configuration with a volume of the opposite sign, which cancels its contribution to the integral.

Indeed, the integration over the independent vectors $\rv$ cannot be unconstrained since they must not cross one another (see, e.g., figure \ref{fig:pack}). To accommodate this condition in two dimensions, we constrain the vectors to rotate always clockwise around particles and anticlockwise around cells. This can be implemented by ensuring that the cross-product of successive vectors must be negative in circulating around particles, and positive in circulating around cells. Thus, defining the `trend' function
\begin{align} \label{eq:trend}
\T(\rv^i, \rv^j) \equiv \rv^i \times \rv^j \ ,
\end{align}
we introduce the following constraints into $Z_a$:
\begin{align} \label{eq:constraints}
\Ttotal \equiv \prod_g^{\rm grains} \prod_{i=1}^{z_g} & \mathcal{H} \left[ - \T ( \rv^{(g, i)}, \rv^{(g, i+1)} ) \right] \nn \\
\cdot \prod_c^{\rm cells} \prod_{i=1}^{z_c} &\mathcal{H} \left[ \T ( \rv^{(c, i)}, \rv^{(c, i+1)} ) \right] \ ,
\end{align}
with $\mathcal{H}$ the Heaviside step function. To satisfy these constraints, we augment the connectivity function by introducing a set of Lagrange multipliers, $\lambda^{(g/c, i)}$:
\begin{align} \label{eq:augment}
\C = \C_0 &- \sum_{g, i} \lambda^{(g, i)} \T ( \rv^{(g, i)}, \rv^{(g, i+1)} ) \nn \\
&+ \sum_{c, i} \lambda^{(c, i)} \T ( \rv^{(c, i)}, \rv^{(c, i+1)} ) \ .
\end{align}

An advantage of this formulation is that it does not complicate the calculation. Our added constraints involve only quadratic terms in $\rhov$ and, therefore, they only modify the numerical values of the entries of the matrix $B$ in the partition function. It follows that our results, eq. (\ref{eq:cexp}-\ref{eq:EOS}) are still valid, but with a modified matrix $B\to B_T$. Using our projection operator, $\J(\T) = \epsilon$ and, hence, $\J(B_T) \ne \id$, resolving the problem of the vanishing mean volume in eq. (\ref{eq:vexp}).

\begin{figure}[h]
\subfloat[]{\includegraphics[clip,width=0.25\textwidth]{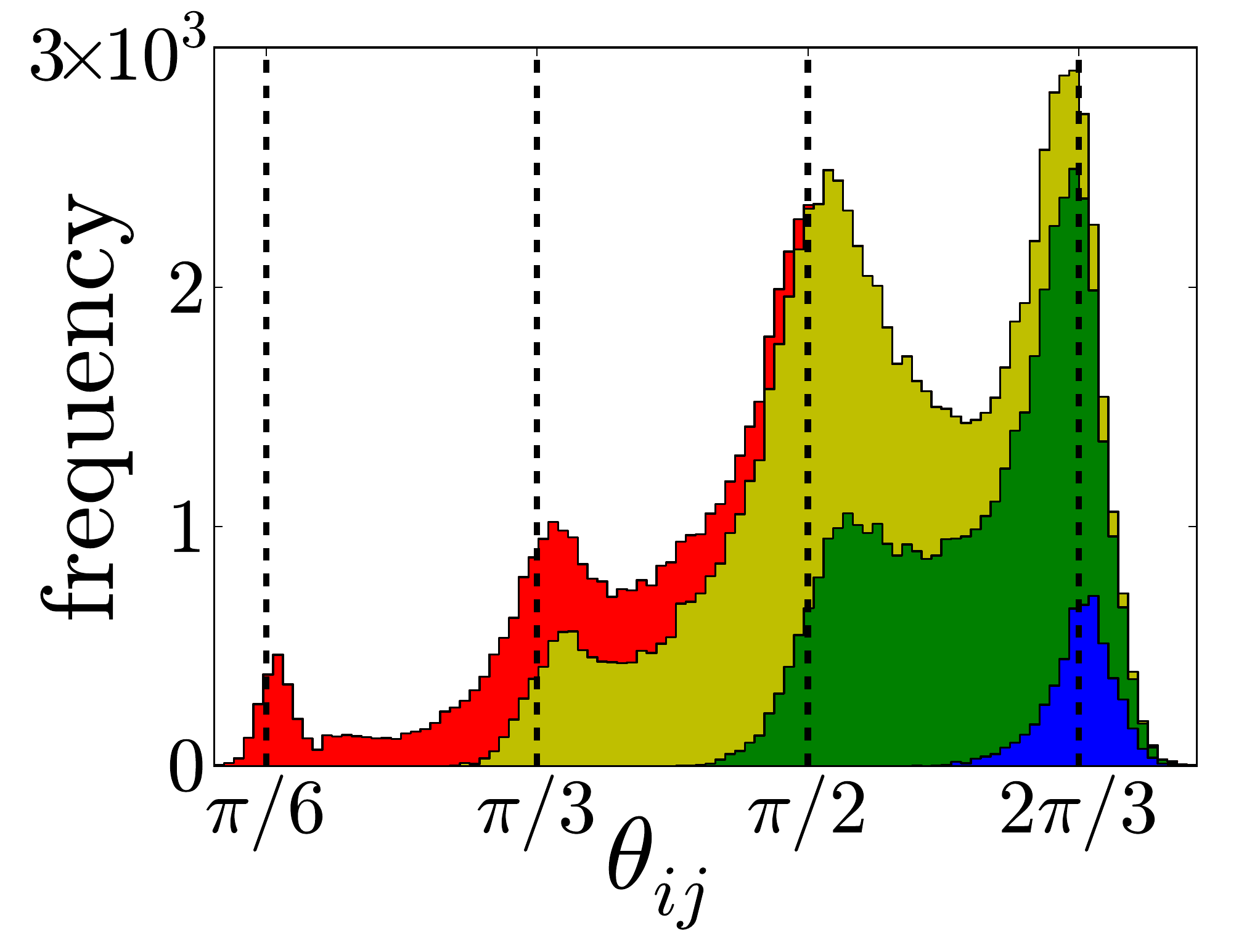} \label{fig:grain_angle}}
\subfloat[]{\includegraphics[clip,width=0.25\textwidth]{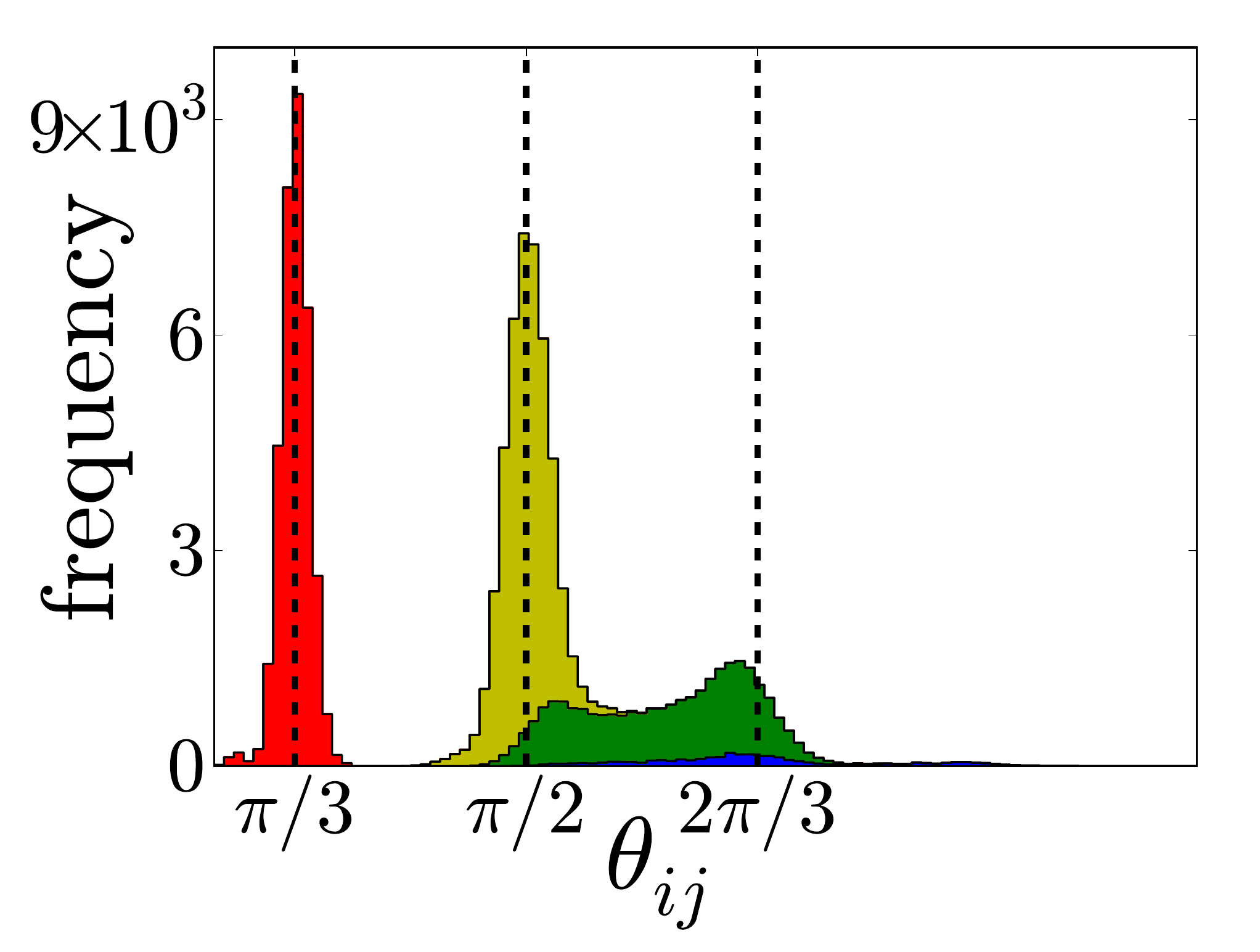} \label{fig:cell_angle}}
\caption{The histograms of $\theta_{ij}$, for 396 systems of 64 particles each, in two cases: (a) when $\rv^i$ and $\rv^j$ share a particle, and (b) when they share a cell. The colours red, yellow, green and blue correspond to a particle/cell of order 3, 4, 5 and 6, respectively. The angle in cells of order-3, for example, is narrowly distributed around $\pi/3$, corresponding to an equilateral triangle. The deviation from $\pi/3$ is due to the distribution of particle sizes.}
\label{fig:angle_distributions}
\end{figure}

We now note that a similar tactic can be employed to take into account correlations between vectors $\rv$, which are inherently present in real granular packs. For example, the angle, $\theta_{ij}$, between two successive vectors, $\rv^i$ and $\rv^j$, assumes very specific values, given the order of the particle or cell that they share. To demonstrate this, we analyse 396 computer-generated systems of 64 soft particles \cite{Maetal16}. The particle radii are normally-distributed, with a mean of 1 and a standard deviation of 0.1 in arbitrary units. The systems were generated at a constant volume fraction, $\phi = 0.84$, and their mean coordination numbers range between 4 and 4.4 (discarding rattlers). Figure \ref{fig:angle_distributions} shows the PDF of $\theta_{ij}$, according to cell and particle order. Given the order of the particle or cell, $\theta_{ij}$ is narrowly distributed around a certain value, $\mu_{ij}^{(\theta)}$, with a certain standard deviation, $\sigma_{ij}^{(\theta)}$. Using a combination of dot-products and cross-products we can construct `potential wells' around these preferred values. Concretely, by adding the following term to the connectivity:
\begin{align} \label{eq:prefer_theta}
\C^{(\theta)}_{ij} &= a (\rv^i \cdot \rv^j) + b (\rv^i \times \rv^j) \nn \\
&= |\rv^i| |\rv^j| (a \cos \theta_{ij} + b \sin \theta_{ij}) \ ,
\end{align}
we create a potential well around $\mu_{ij}^{(\theta)} = \arctan(b/a)$. The magnitudes of $a$ and $b$ determine the depth of the well and thus determine $\sigma_{ij}^{(\theta)}$. Such a term can be added for every two successive vectors, with the appropriate parameters $a$ and $b$. These additions, like the augmented connectivity of eq. (\ref{eq:augment}), are all quadratic in $\rhov$, and thus do not complicate the calculation. \\

The addition of such terms introduces explicitly interactions between the DFs, a concept that has been absent from the original formulation of GSM.

\begin{figure}[h]
\subfloat[]{\includegraphics[clip,width=0.25\textwidth]{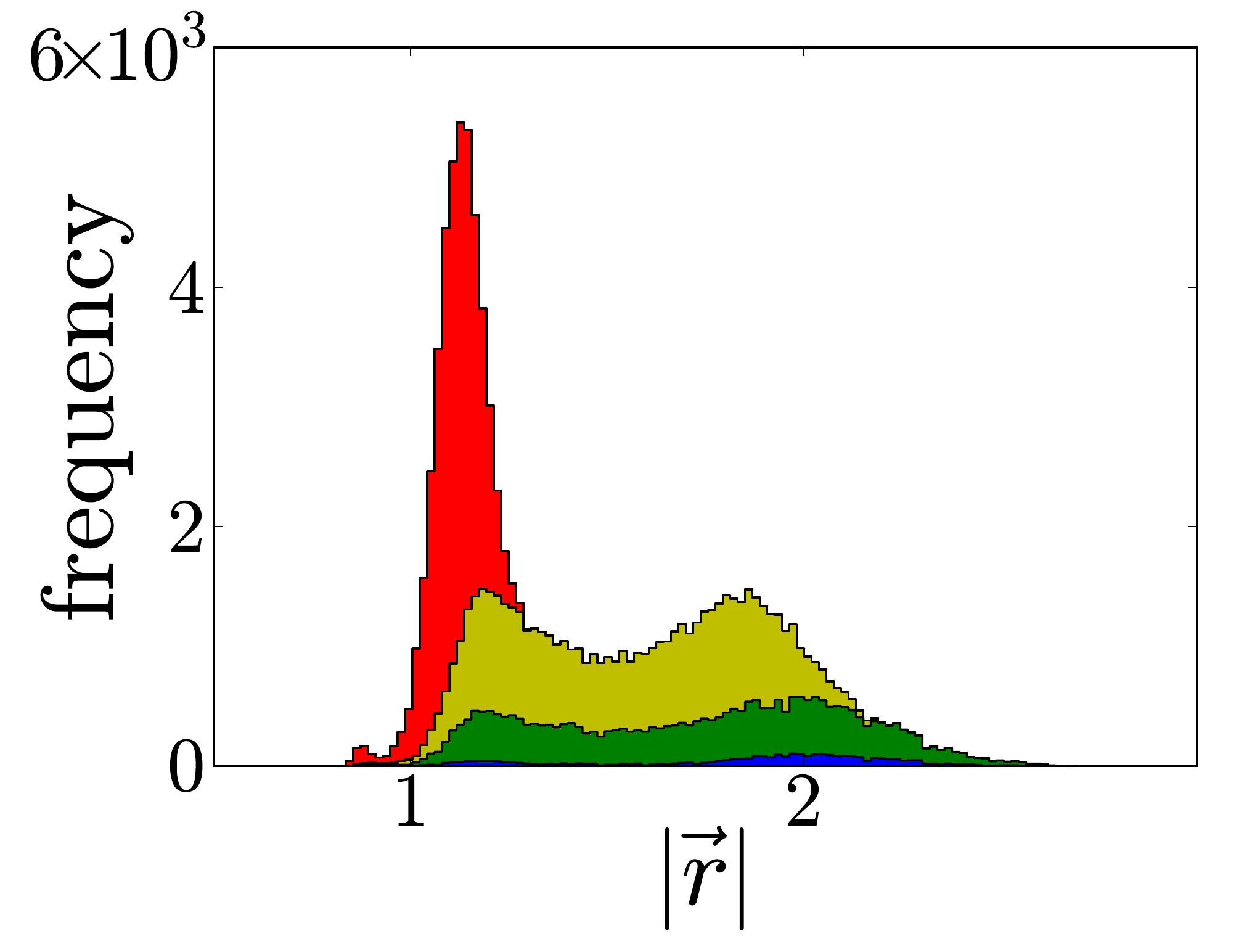} \label{fig:r_length_by_cell}}
\subfloat[]{\includegraphics[clip,width=0.25\textwidth]{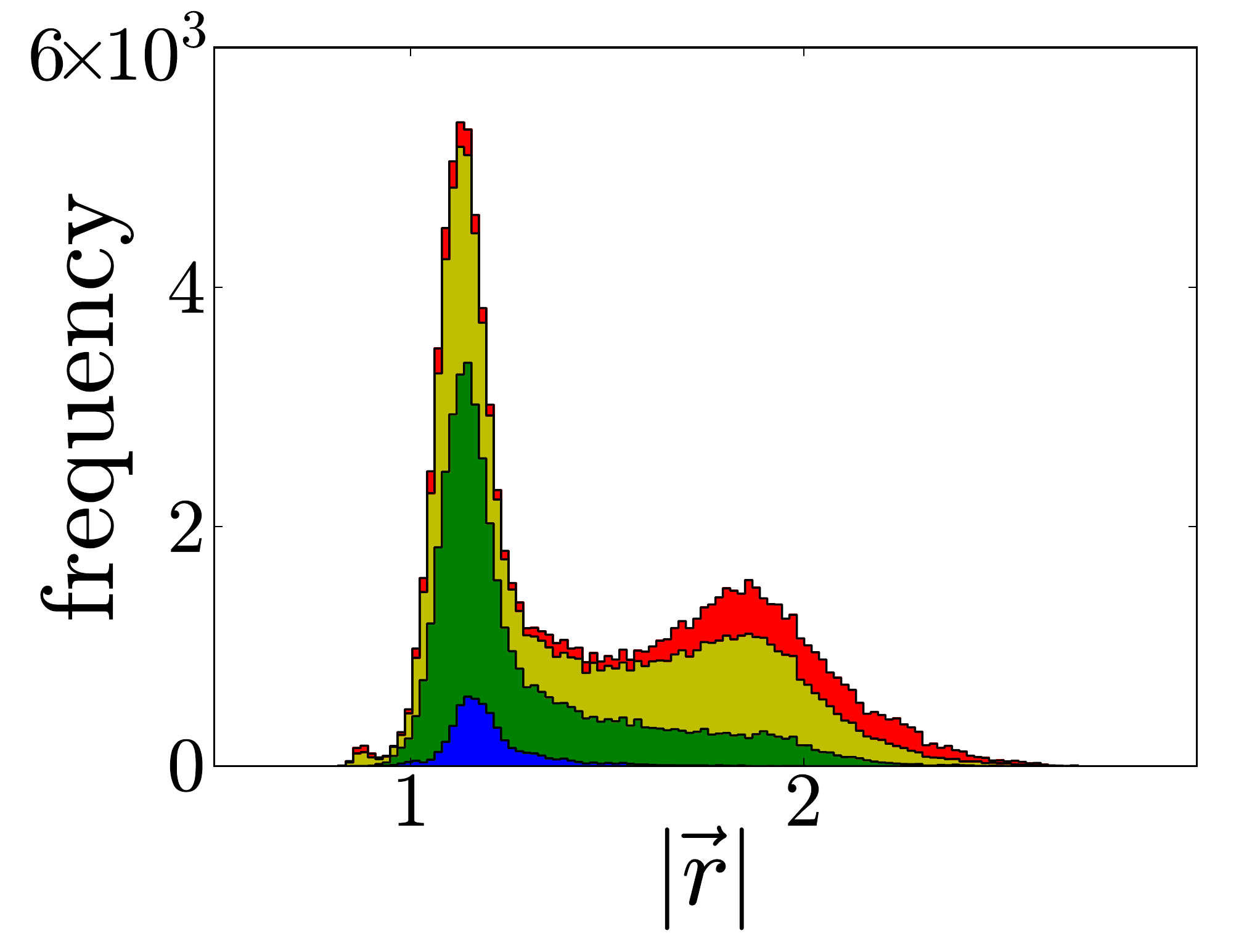} \label{fig:r_length_by_grain}} \\
\subfloat[]{\includegraphics[clip,width=0.25\textwidth]{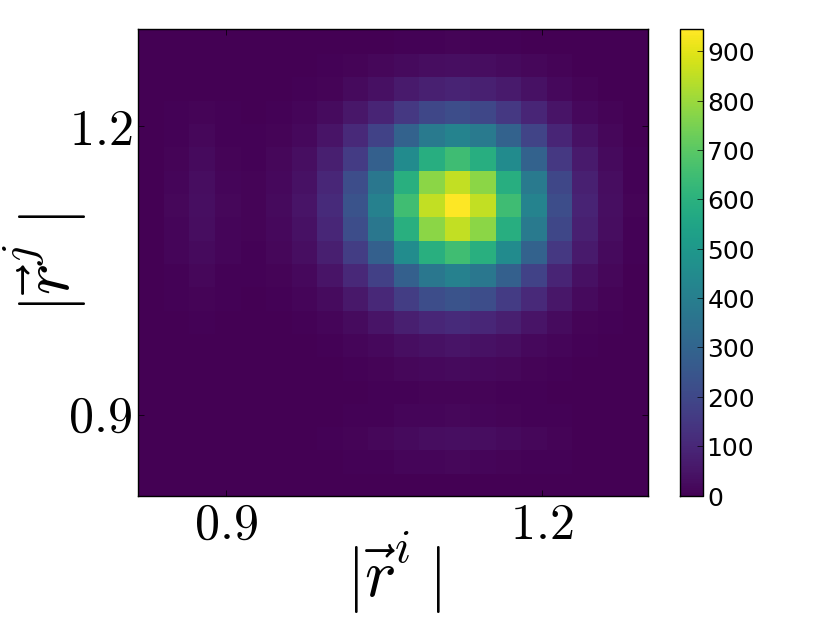} \label{fig:r_length3_expected}}
\subfloat[]{\includegraphics[clip,width=0.25\textwidth]{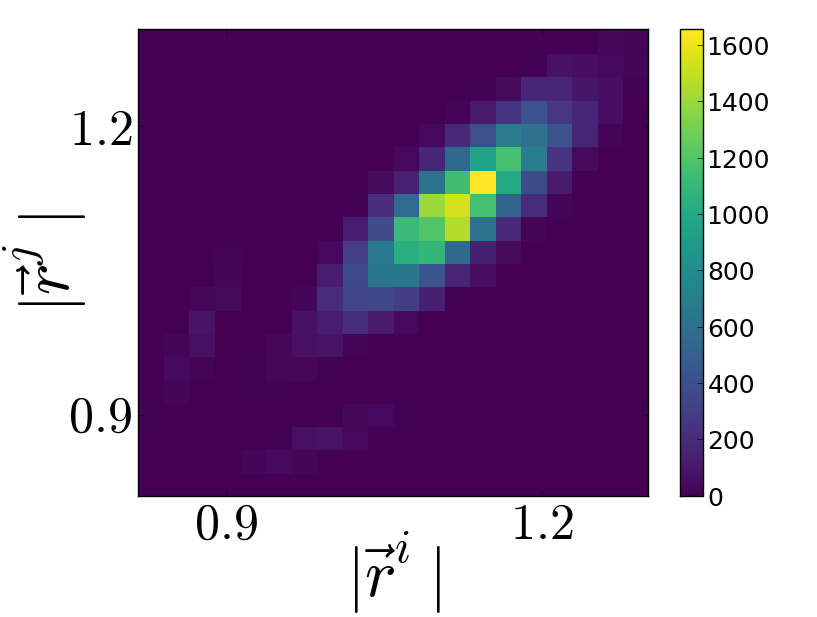} \label{fig:r_length3_actual}} \\
\subfloat[]{\includegraphics[clip,width=0.25\textwidth]{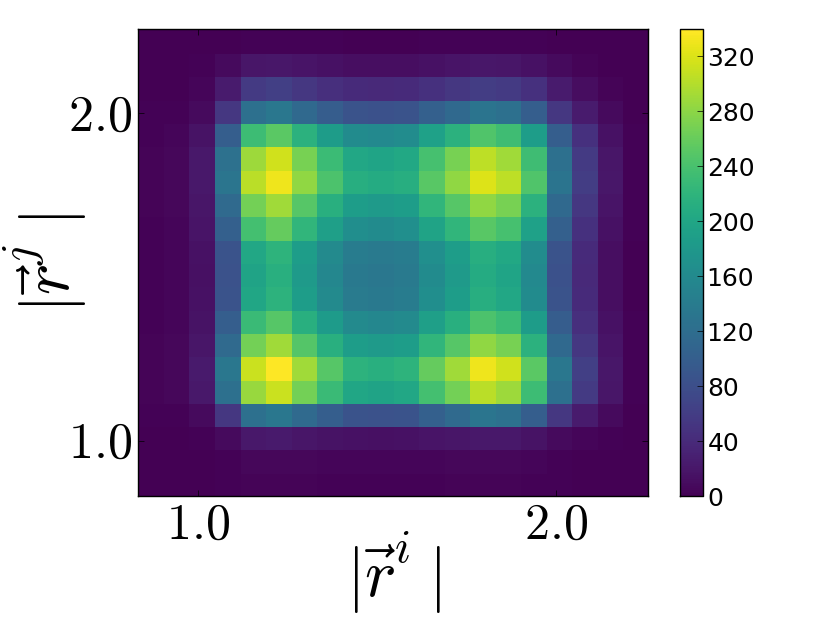} \label{fig:r_length4_expected}}
\subfloat[]{\includegraphics[clip,width=0.25\textwidth]{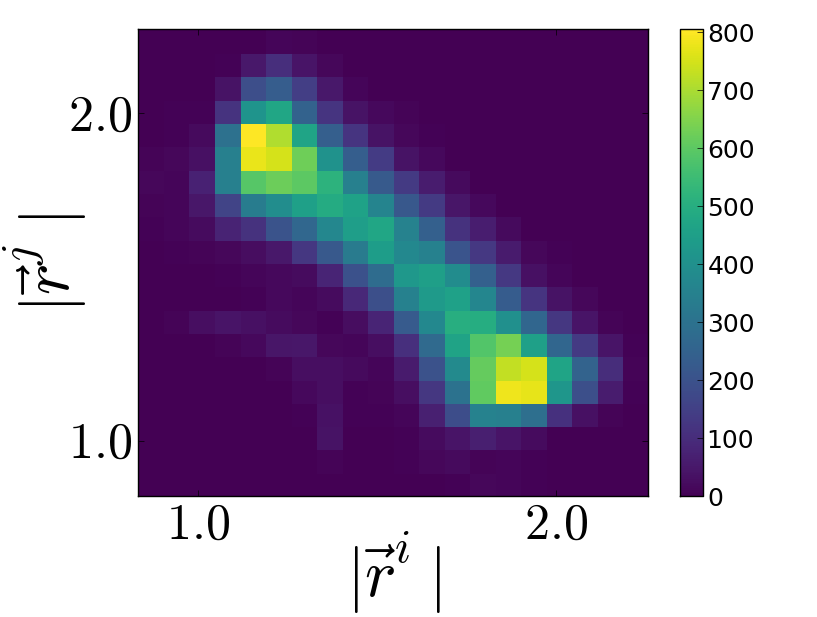} \label{fig:r_length4_actual}}
\caption{(a, b) The histogram of $|\rv|$, for 396 systems of 64 particles each, in units of the average particle radius. Each vector $\rv$ belongs to one particle and one cell, and the histogram is divided according to the order of the cell (a) and the particle (b), with orders 3, 4, 5 and 6 represented by red, yellow, green and blue, respectively. (c, e) The expected 2D histogram of successive vectors, $\rv^i$ and $\rv^j$, around 3- and 4-order cells, respectively, according to (a), and not taking correlations into account. (d, f) The actual 2D histogram corresponding to (c) and (e). The heat maps on the right differ from those on the left due to positive correlations for order-3 cells and negative for order-4 cells.}
\label{fig:r_length_distributions}
\end{figure}

Another example of an interaction between DFs is the inherent correlation between the lengths of successive vectors $\rv$ around a particle. Figures \ref{fig:r_length_by_cell} and \ref{fig:r_length_by_grain} show the histogram of $|\rv|$, sampled in two different ways: one by the order of its cell (figure \ref{fig:r_length_by_cell}) and the other by the order of its particle (figure \ref{fig:r_length_by_grain}). We see that $|\rv|$ increases with cell order and decreases with particle order. This has a geometrical origin -- the more vectors surround a particle, the shorter they must be; but the more vectors constitute a cell, the longer they can be (see figure \ref{fig:pack}). Let us focus on order-3 cells -- the tall peak in figure \ref{fig:r_length_by_cell}. Figures \ref{fig:r_length3_expected} and \ref{fig:r_length3_actual} show, respectively, 2D histograms (shown as heat maps) of the expected correlation-free and actual occurrence frequencies of the lengths of successive vectors, $\rv^i$ and $\rv^j$, around order-3 cells. The differences between the two indicate a positive correlation between the lengths, which is the result of a correlation in the lengths of the three vectors upon increase of the size of any of the 3 particles around the cell.

The same can be done for particles of any number of contacts or cells of any order. In figures \ref{fig:r_length4_expected} and \ref{fig:r_length4_actual} we show, respectively, the expected correlation-free and actual distributions for order-4 cells. Here too the two differ significantly, but with a negative correlation that arises from a geometrical origin: the lengths of two opposite edges in order-4 cells are sensitive to the distance between the two respective particles (see figure \ref{fig:pack}), the longer one pair of opposite edges, the shorter the other.

To take account of such correlations, we can include in the connectivity function another `interaction-like' term between two successive vectors $\rv^i$ and $\rv^j$:
\begin{align} \label{eq:prefer_theta}
\C^{|r|}_{ij} &= \left( |\rv^i|^2 - |\rv^j|^2 \right)^2 \ ,
\end{align}
which is quartic in $\rhov$. The calculation of the partition function with quartic terms is more complex and is left for a future study.

\section{The topological entropy} \label{sec:topological}

The topological entropy arises from all the possible changes in the contact network, i.e. the topology. Each such change, which corresponds to making or breaking a contact, effects a change in the matrix $A$. The inverse, however, is not generally true - many different matrices $A$ may correspond to the same topology because the choice of the spanning tree is not unique. Thus, to obtain the number of possible topologies we need to divide the total number of possible matrices $A$, $\Omega_t$, by the number of spanning trees describing a typical configuration, $\Nst$. We call the latter multiplicity of a microstate.

The statistics of $A$ can be described by a multi-variate probability density function of its components, $P(\{A_{ij}\})$. The partition function is then
\begin{align} \label{eq:z_with_topological}
Z = \sum_{A_{ij}} P(\{A_{ij}\}) \cdot Z_a(A) \ ,
\end{align}
where the sum is over all possible values of $A_{ij}$.

To estimate $P(\{A_{ij}\})$, we first recall that the top part of $A$, corresponding to the $N_\rho$ independent vectors, is always a unit matrix. The rest of the matrix consists of entries of $0$ and $\pm 1$, whose $ij$ positions can change from topology to topology. 
A straightforward naive approximation is to assume that each $A_{ij}$ is independent and can take the values $0$, $1$ and $-1$ with respective probabilities $P_0$, $P_1$ and $P_{-1}=1-P_0-P_1$. These probabilities are constrained by the requirement that $P_1 + P_{-1} = \lavg / N_\rho$, where $\lavg$ is the average number of steps along the spanning tree, needed to describe a dependent vector $\rv$ in terms of the independent ones. The average $\lavg$ is carried out over all possible topologies (possible matrices $A$) generated by the same packing procedure of a specific collection of $N$ particles, resulting in the same mean coordination number $\bar{z}$.

However, this estimate can be improved. Since each line of $A$ represents a route along the spanning tree, a more realistic estimate is
\begin{align}
P(\{A_{ij}\}) = \prod_{i} P(n^{(i)}_0, n^{(i)}_1, n^{(i)}_{-1}) \ , \nn
\end{align}
with $n^{(i)}_0$, $n^{(i)}_1$ and $n^{(i)}_{-1}$, respectively, the number of $0$'s, $1$'s and $-1$'s in the $i$'th row of $A$.
Different topologies correspond to different bottom parts of A, consisting of $N_r - N_\rho \sim 2 N_c - N_c = N_c$ rows and $N_\rho \sim N_c$ columns. Typically, each of those $N_c$ rows has $n \equiv \lavg / 2$, of $1$s and $-1$s. Therefore, with $n \ll N_c$, the total number of possible different $A$ matrices is
\begin{align} \label{eq:omega_tot}
\Omega_t \approx \left[ {N_c \choose n} {N_c \choose n} \right]^{N_c} \approx \left( \frac{N_c^n}{n!} \right)^{2 N_c} \ ,
\end{align}
and we have
\begin{align}
\ln \Omega_t = 2 N_c \left[ n \ln N_c - \ln n! \right] + O(n) \ .
\end{align}

We now need to estimate the multiplicity of spanning trees per topology, $\Nst$. This can be done using a result from graph theory: for a wide class of graphs \cite{Ly2005},
\begin{align} \label{eq:zetaG}
\lim_{N_V \to \infty}\Nst^{(G)} = e^{\zeta_G N_V} \ ,
\end{align}
where $\zeta_G$ is a constant, whose value depends on the particular topology of the graph $G$, and $N_V$ is the number of the graph's vertices, which is $N_c$ in our case. A straightforward upper bound on $\Nst$ can be established by assuming that any choice of $N_\rho (\sim N_c)$ vectors out of the possible $N_r (\sim 2 N_c)$ is a possible spanning tree,
\begin{align}
\Nst < {{2 N_c} \choose {N_c}} \approx \sqrt{\frac{4 N_c}{\pi}} 2^{2 N_c} \ .
\end{align}
Using this bound we get, in the large $N_c$ limit,
\begin{align} \label{eq:upper_bound}
\zeta_G < 1.386 + O(\ln{N_c}/N_c) \ .
\end{align}
This bound is too high since most choices of a random set of $N_c$ vectors $\rv$ include forbidden vector combinations of closed loops. 
A better estimate would be by using a known bound for regular graphs \cite{Mc83, ShWu00}:
\begin{align} \label{eq:KG}
\Nst^{(G)} \leq \frac{2\ln N_V}{N_V K \ln K} (C_K)^{N_V} \ ,
\end{align}
where $K$ is the valency of the vertices of $G$, and $C_K \equiv [K - 1]^{K - 1} / [K (K - 2)]^{K / 2 - 1}$. Thus, in the large $N_V$ limit, $\zeta_G \le \ln C_K$. Almost all the vertices of graphs describing contact network connect four vectors $\rv$ (see, e.g., figure \ref{fig:pack}) and have $K=4$. The exceptions are: (i) $K=3$ for contacts of two-contact particles, but these consist a small fraction, $\varepsilon$, of all particles; (ii) $K=2$ for chains of two-contact particles, but these are unstable mechanically and rarer still; (iii) $K=2$ for generic boundary contacts, but the number of these scale as $M \ll N_c$. The total number of contacts is, $N_c = (N \bar{z} + M) / 2$ and the total number of edges is $N_E = N(\bar{z} - \varepsilon)$. Since $M \ll N$ and $\varepsilon \ll 1$ for sufficiently large systems,
\begin{align} \label{eq:Kz}
K = \frac{2N_E}{N_V} \approx 4 \left( 1 - \frac{M}{N \bar{z}} - \frac{\varepsilon}{\bar{z}} \right) \approx 4 \ .
\end{align}
Using eq. (\ref{eq:KG}), we then obtain $C_K = (1.5)^3$ and an improved upper bound, compared to (\ref{eq:upper_bound}),
\begin{align} \label{eq:regular_upper_bound}
\Nst \leq (1.5)^{3N_c} \hs , \hs \zeta_G \le 1.216 \ .
\end{align}

However, $\zeta_G$ depends not only on $K$ but also on further details of the structure. For example,  the square and Kagom\'e lattices, both regular graphs with $K=4$, have different values, $\zeta_{\rm Sq} = 1.166$ and $\zeta_{\rm Kag} = 1.136$ \cite{Mc83, ShWu00}. The difference between these structures is their distributions of cell orders, or the $K$ value of their dual graphs.  The dual of the square lattice is also regular with $K=4$, but the dual of the Kagom\'e lattice has $2/3$ vertices with $K=3$ and $1/3$ with $K=6$. Since, of the two graphs, the latter's distribution of cell orders is closer to that of real systems, in which these range between 3 and 6 (see figure \ref{fig:pack}), then we conjecture that the  Kagom\'e lattice should describe our systems more closely. This provides us the best estimate,
\begin{align} \label{eq:NstN}
\zeta_G \approx 1.136 \ ,
\end{align}
which is lower than both eq. (\ref{eq:upper_bound}) and (\ref{eq:regular_upper_bound}).

\begin{figure}[h]
\subfloat[]{\includegraphics[clip,width=0.25\textwidth]{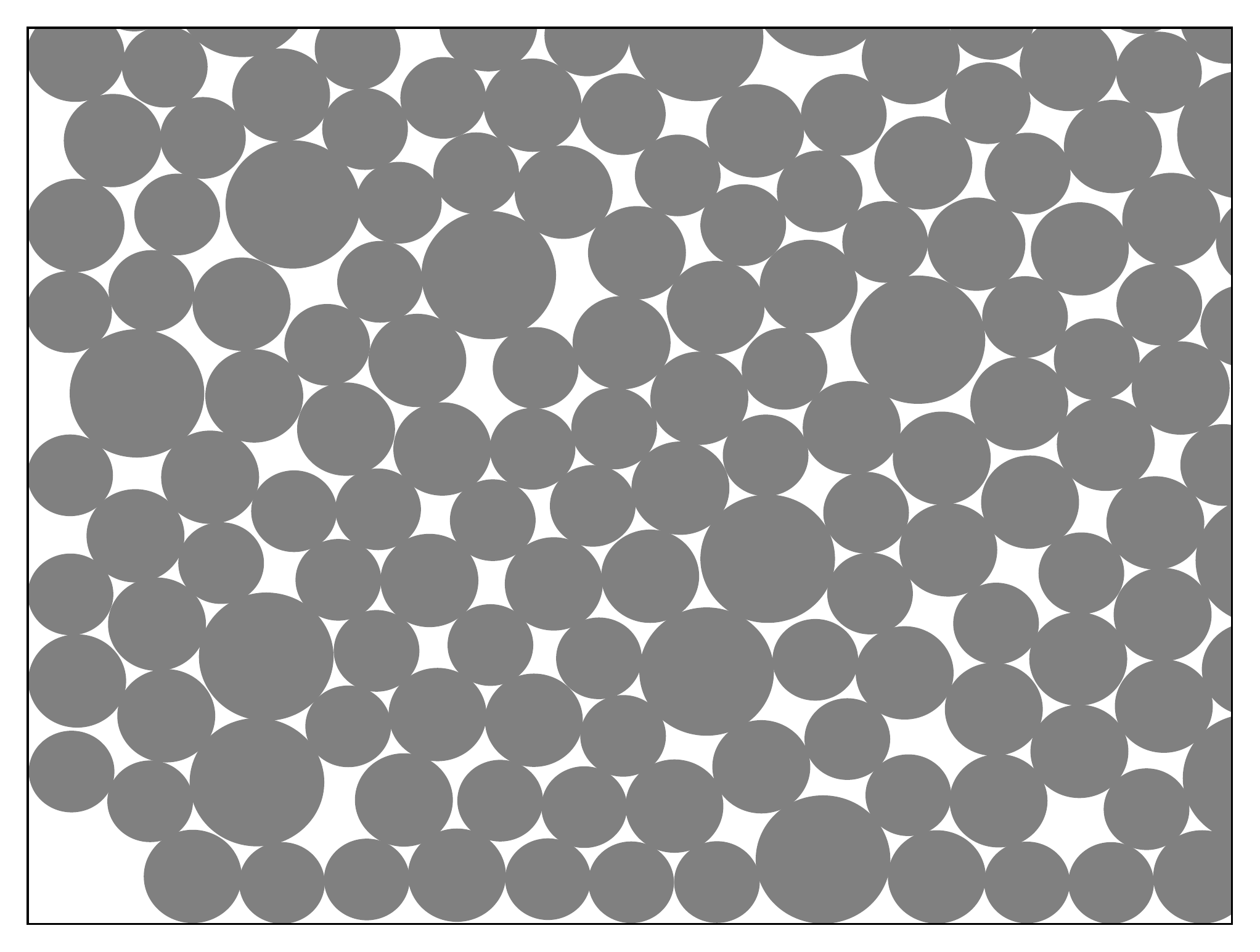}}
\subfloat[]{\includegraphics[clip,width=0.25\textwidth]{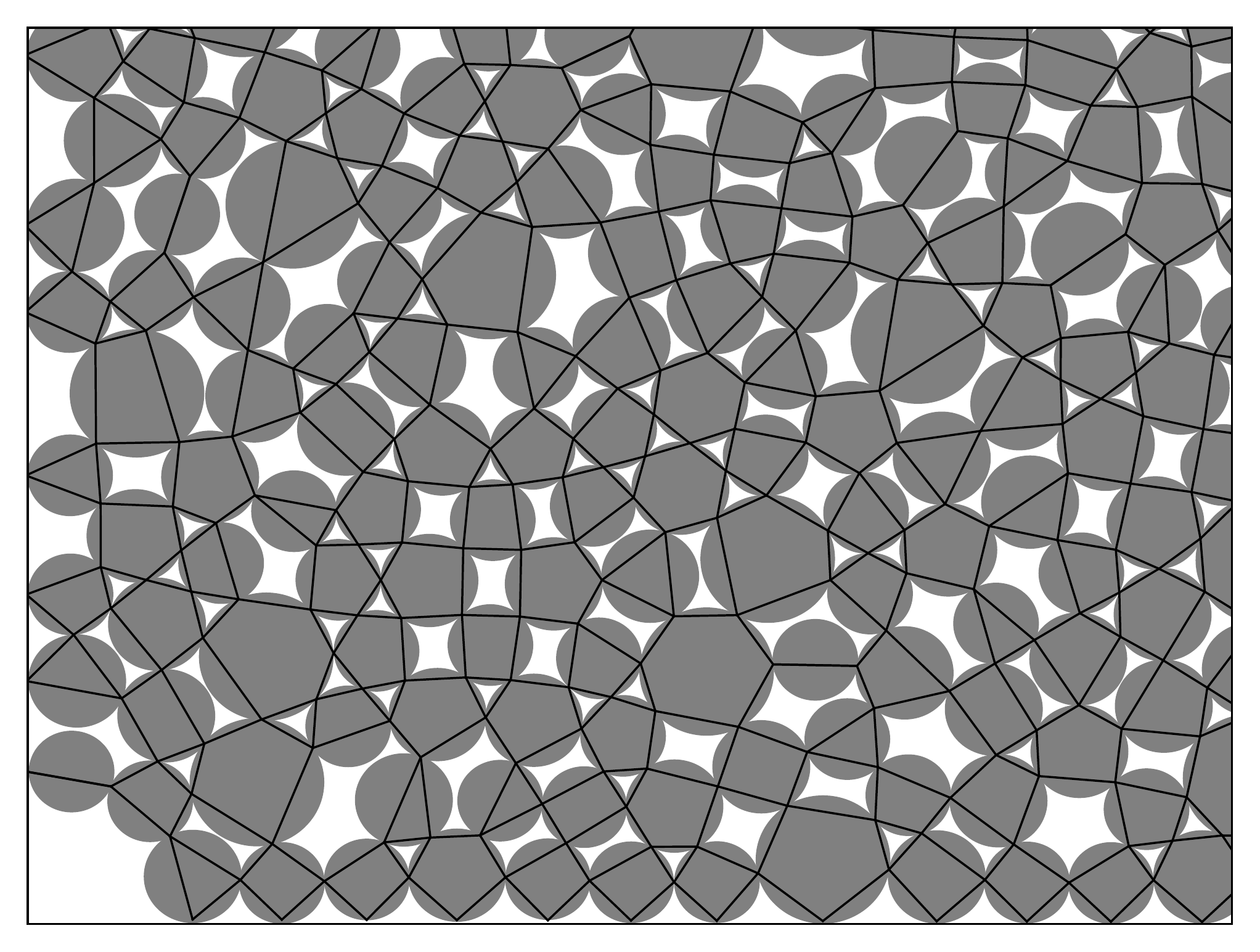}} \\
\subfloat[]{\includegraphics[clip,width=0.25\textwidth]{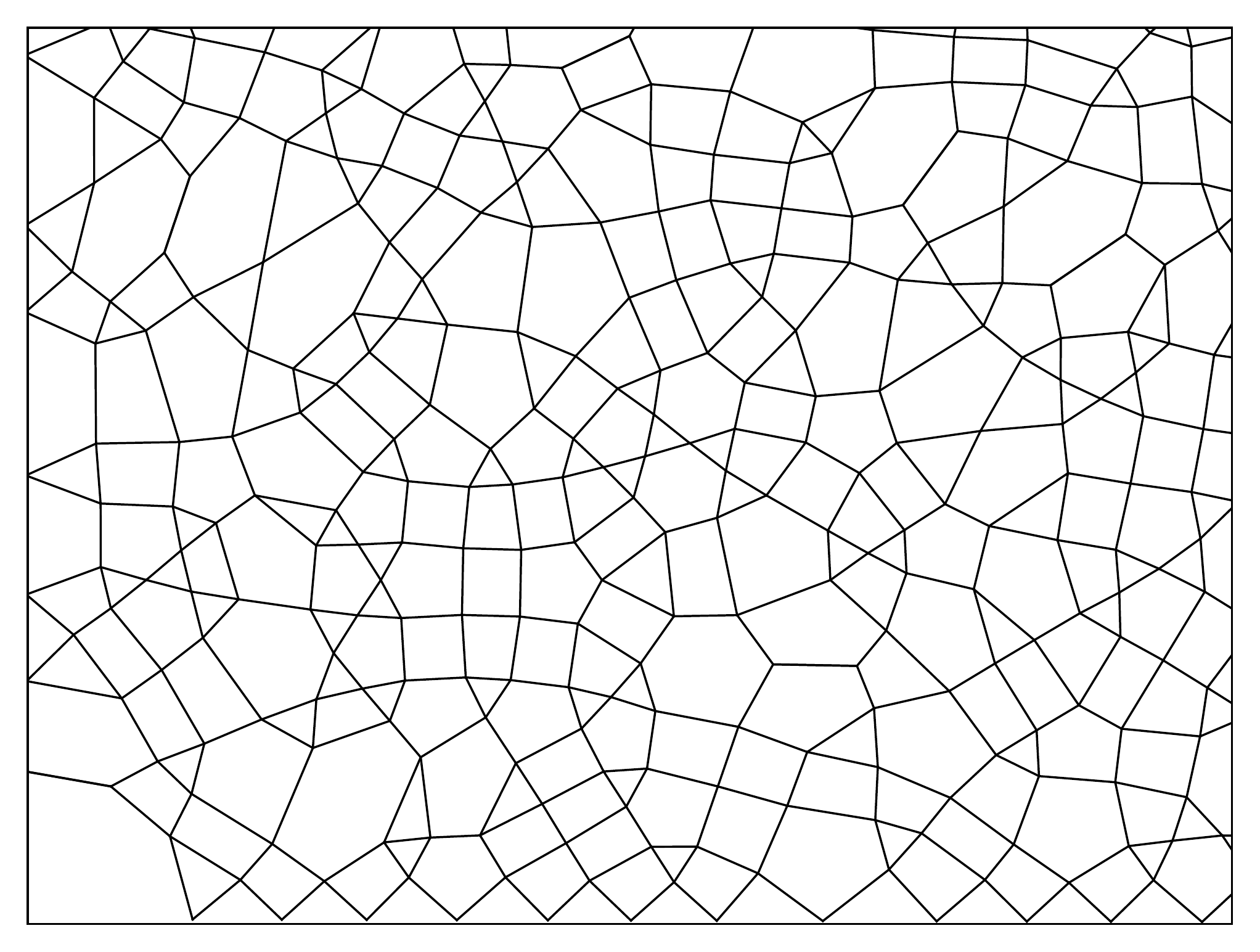}}
\subfloat[]{\includegraphics[clip,width=0.25\textwidth]{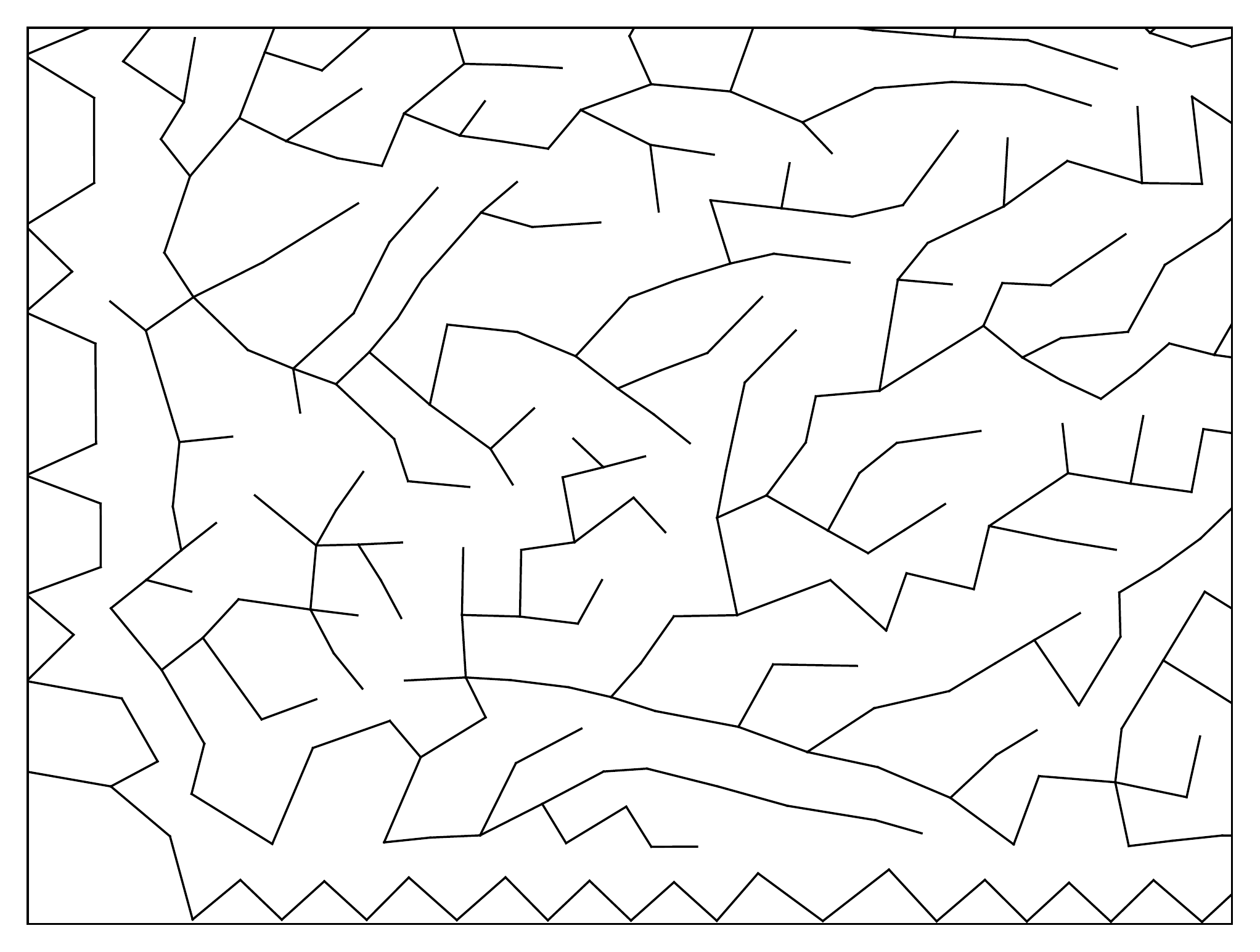}}
\caption{(a) The bottom-left corner of a tri-disperse 2D experimental granular system, produced by the 3SR Lab \cite{Caetal97, *Rietal12}. (b) The network of vectors $\rv$, circulating every particle and cell. The directionality of the vectors, clockwise around particles and anti-clockwise around cells, is not plotted to avoid cluttering. (c) The connectivity network of this system. (d) One choice of a spanning tree, i.e. a subset of independent vectors $\rv$. All boundary vectors $\rv$ but one are chosen.}
\label{fig:gael}
\end{figure}

\begin{figure}[h]
\includegraphics[width=0.5\textwidth]{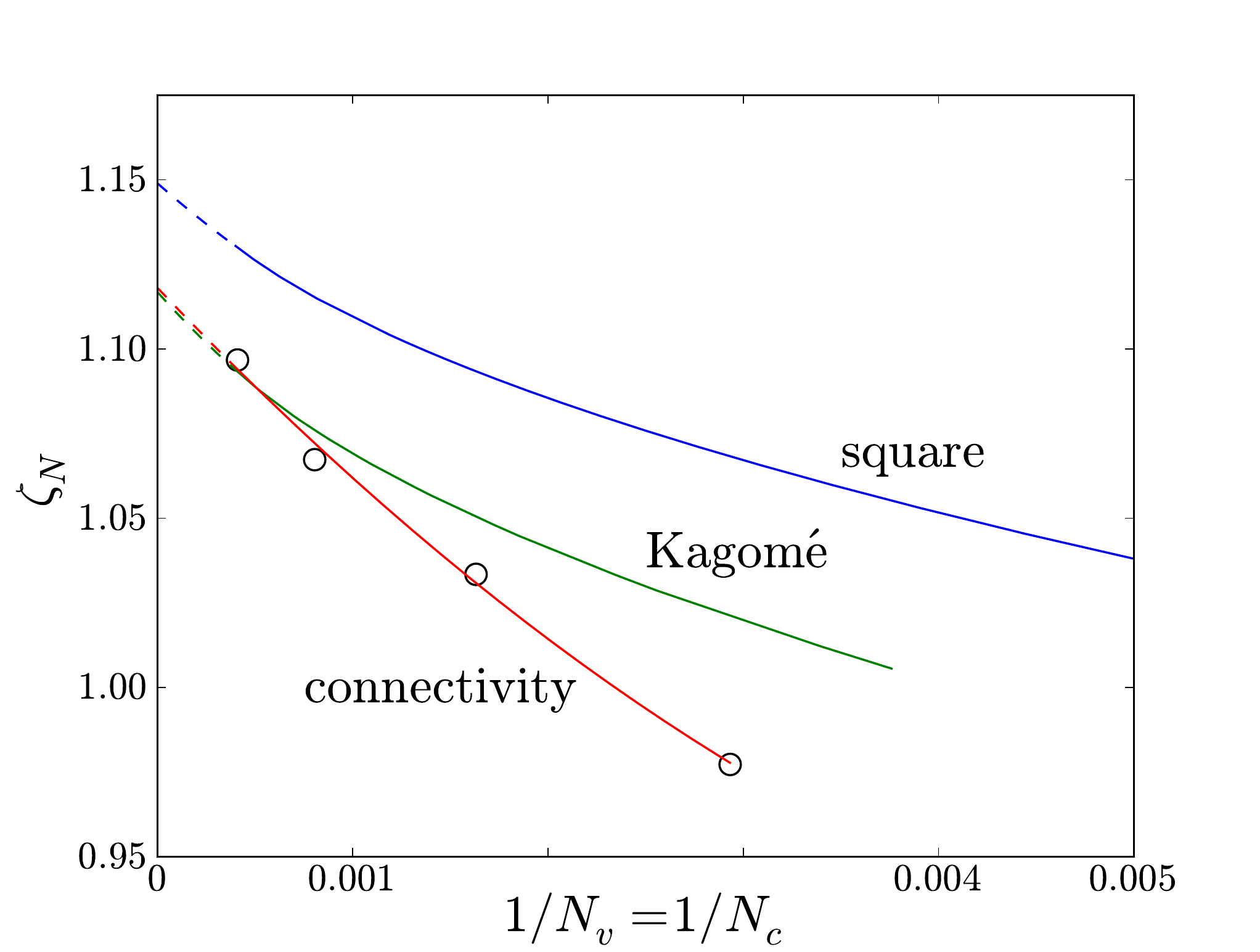}
\caption{$\zeta_N$, defined in eq. (\ref{eq:zetaG}), as a function of the inverse of the number of vertices, $1/N_V$, for different graphs: the Kagom\'e lattice (green line), the square lattice (blue line) and the connectivity network of the experimental 2D systems (black open circles, interpolated by red line). The value of $\zeta_N$ for the Kagom\'e lattice and the experimental systems is calculated numerically, by constructing the graphs and using Kirchhoff's theorem to get the number of spanning trees. $\zeta_N$ for the square lattice is calculated analytically. In dashed lines are quadratic extrapolations to $1/N_V \to 0$.}
\label{fig:nst}
\end{figure}

To test this result, we analysed 2D experimental systems, each of 1172 discs of three different radii, produced by the 3SR Lab, as described in \cite{Caetal97, *Rietal12}. We divided the image of each system to several non-overlapping sub-systems of different sizes, and constructed the contact network for each sub-system. Figure \ref{fig:gael} demonstrates this procedure and shows one choice of a spanning tree. We counted the number of possible spanning trees for each sub-system using Kirchhoff's theorem \cite{Ki1847}. We then calculated $\zeta_G$ as a function of system size, using eq. (\ref{eq:zetaG}). Figure \ref{fig:nst} shows $\zeta_G$ as a function of $1/N_c$ alongside $\zeta_{\rm Sq}$ and $\zeta_{\rm Kag}$ of lattices of the same size, for which the values were also obtained by using Kirchhoff's theorem. Indeed, the value of $\zeta_G$ in the disordered systems converges to that of the Kagom\'e lattice, supporting our result (\ref{eq:NstN}).

\begin{figure}[h]
\includegraphics[width=0.5\textwidth]{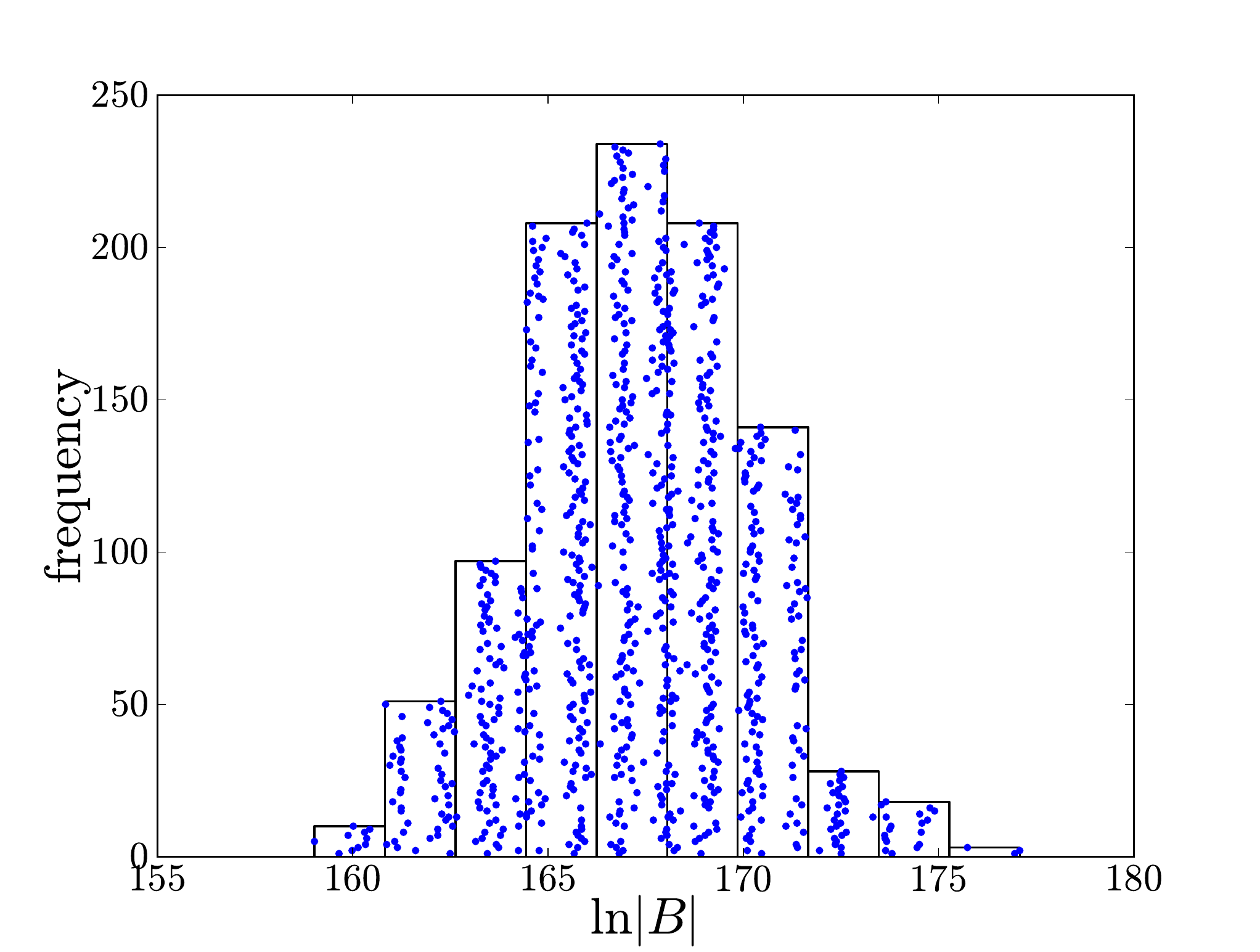}
\caption{The histogram of $\ln |B|$ for 1000 systems of 64 particles each (white bars). Each blue dot represents a single system, at the appropriate $x$-value and a random $y$-value.}
\label{fig:logB}
\end{figure}

Having a reliable estimate of the multiplicity, $\Nst$, we can now estimate the topological entropy:
\begin{align} \label{eq:topological_entropy_orig}
S_t &= \ln (\Omega_t / \Nst) \nn \\
&= 2 n N_c \ln N_c - \left(2 \ln n! + \zeta_G \right) N_c + O(n) \ .
\end{align}
In the calculation above we assumed that $S_a (A)$ depends only weakly on the topology. To test this assumption, we first note that the contribution of the boundary forces to $S_a$ is negligible compared to the structural configurations and that $S_a \approx \ln{|B|} + O(\sqrt{N})$. Calculating the value of $\ln{|B|}$ for $1000$ systems of $64$ soft disks each, computer-generated by the same protocol \cite{Maetal16}, we plot its distribution in Figure \ref{fig:logB}. Indeed, we find that $\ln{|B|}$ has a well-defined value with a relative width of $2.9/167=1.7\%$, supporting well the above assumption.
Using $N_c = N \bar{z} / 2$ and $n = \lavg / 2$, we can express $S_t$ in terms of the number of particles, $N$:
\begin{align} \label{eq:topological_entropy}
S_t = \frac{\bar{z}}{2} \lavg N \ln N + \frac{\bar{z}}{2} \left( \lavg \ln \frac{e \bar{z}}{\lavg} - \zeta_G \right) N \ .
\end{align}
Thus, $S_t$ is dominated by the term $N \ln N$, reflecting the fact that it corresponds to arrangements of all the particles in a network, whose number is of order $N!$.

This result is very significant -- it indicates that the topological structural entropy is extensive, i.e. linear in system size, only when subtracting $(\bar{z}\lavg / 2) \ln{N!}$ from it. This observation is reminiscent of the much discussed $\ln{N!}$ subtraction in thermal statistical mechanics. However, it is interesting that, in GSM, only the topological entropy incurs this term. Thus, together with result (\ref{eq:AffEntropy}), the total structural entropy, $S_a + S_t$, is extensive. Significantly, relation (\ref{eq:topological_entropy}) is supported by, and provides the theoretical explanation for, the numerical observations in \cite{Asetal14, Maetal16}, who observed directly the $N!$ factor. 

\section{Discussion} \label{sec:discussion}

In this paper, we extended the connectivity-based granular statistical mechanics, proposed in \cite{Bletal16}, and derived a number of fundamental results as follows. 
Firstly, we calculated the complete partition function, eq. (\ref{eq:z_orig}), beyond the structural contribution and included the explicit contribution of the stress ensemble. The quadratic exponential in eq. (\ref{eq:z_simple}) allowed us to not only solve it, eq. (\ref{eq:z_final}), but also to calculate explicitly expectation values for: (i) the connectivity, $\Cexp_a$, showing that there exists an equipartition principle for it; (ii) the squared-norm of the force vector, $\gexp_a$; (iii) the boundary normal stress, $\sexp_a$; and (iv) the volume, $\Vexp_a$. Using these, we derived {\it a new equation of state}, eq. (\ref{eq:EOS}).

Secondly, we identified two main sources for the structural entropy, {\it affine and topological}, with the former describing microstates with the same topology of the contact network and the latter describing microstates of different topologies. The connectivity-based formulation was shown to be convenient for separating these two contributions, eq. (\ref{eq:z_with_topological}), with all the topology encoded in the connectivity matrix $A$. This separation made it possible to calculate explicitly each of these contributions. We calculated the affine entropy explicitly from eq. (\ref{eq:entropy}) and found that it scales linearly with the number of particles, i.e. it is extensive.

Thirdly, we established that, to calculate the affine contribution to the partition function, one must take into consideration explicitly the correlations between the DFs as constraints and that ignoring these constraints lead to grossly unphysical results, such as a vanishing mean volume. We then modified the connectivity function to include these constraints, eq. (\ref{eq:augment}), and showed that these remedy the calculation and give more physical results.

Fourthly, we showed that our method of including constraints can be used to describe correlations between the DFs as {\it interactions}. This constitutes an extension of the granular statistical mechanics formalism beyond the traditional analogue of self-energy-like description. We demonstrated that granular systems possess inherent positive and negative correlations between angles, as well as lengths, of successive DFs along connectivity loops and outlined how to include these as interaction terms in the modified connectivity function. These interactions are convenient in that they resemble nearest neighbour interactions in more traditional systems.

Fifthly, we calculated the topological entropy, using the statistics of the matrix $A$, which describes the topology of a granular assembly as a spanning tree \cite{Jo14}. This calculation was complicated by the concept of {\it multiplicity}, namely, that a specific structural configuration can be described by $N_{ST}$ spanning trees. We calculated $N_{ST}\sim e^{1.136 \bar{z} N / 2}$, yielding that the number of microstates is $\Omega_t / N_{ST}$. It follows that the overall entropy is extensive {\it only when subtracting} $(\bar{z} \lavg / 2) \ln{N!}$ from it. This result provides a theoretical explanation of the observations in \cite{Asetal14, Maetal16}.

Sixthly, by calculating explicitly the structure and forces based entropies, we established that, in the large system limit, {\it the latter is negligible relative to the former}. This result is the direct consequence that the structure phase space is of size $N$ while the force phase space is of size $N^{(d - 1) / d}$. Another consequence is that the dependence of expectation values of structural properties on the angoricity tensor \cite{EdBl05} is negligible, at least for rigid particles. The implication of this conclusion for forces-based expectation values remains to be studied, but we showed that the equation of state (\ref{eq:EOS}), which includes the boundary stress, depends negligibly on the angoricity unless its value is extremely small.

Our results hold for an ensemble of mechanically stable granular packs, all generated by the same protocol and all having the same mean coordination number, $\bar{z}$. Extending the analysis to distributions of $\bar{z}$ across the ensemble is possible, in principle, albeit cumbersome, as it involves enumeration over matrices $A$ of varying dimensions. We reiterating that our estimate of the topological entropy, $S_t$, was made under the assumption, supported numerically for a certain class of systems in section \ref{sec:topological}, that different topologies have similar occurrence probabilities. When this assumption does not hold, eq. (\ref{eq:topological_entropy_orig}) must be replaced by the more general Gibbs entropy, $S_t = - \sum P_t \ln P_t$, with $P_t$ the probability of each topology. Thus, eq. (\ref{eq:topological_entropy_orig}) is an upper bound since a uniform distribution maximises the entropy.

The formalism presented here can be extended readily to 3D systems, following the same conceptual approach. First, one constructs a network of $\rv$-vectors that form a convex hull around each grain \cite{BlEd06, Fretal08}. Then, choosing a spanning tree of this network, using the same principle as described above, the connectivity function can be calculated straightforwardly. More care is required in calculating the stress ensemble, since 3D cells are surrounded by both grains and throats, which are the openings connecting neighbouring cells. Each throat is made of a loop of $\rv$-vectors. As in 2D, the intergranular forces can be solved for in terms of the boundary forces and, if required, additional constitutive relations. The boundary forces are the DFs of the stress ensemble. It then follows that one can define the 3D equivalents of the matrices $E$ and $H$ and the rest of the analysis is the same as above.

We look forward to numerical and experimental tests of the new formulation.

\section{Acknowledgements}

SA acknowledges financial support by the Alan Howard Scholarship, and thanks S. Martiniani for providing the computer-generated systems \cite{Maetal16}.

\patchcmd{\appendices}{\quad}{. \ \ }{}{}
\numberwithin{equation}{section}
\patchcmd{\theequation}{.}{}{}{}

\begin{appendices}

\section{Calculating expectation values and entropy} \label{appendix:calc}

The different expectation values are all derivatives of $\ln Z_a$ and, using eq. (\ref{eq:z_final}), it is convenient to define
\begin{align}
\ln Z_a = \ln Z_0 + \ln Z_1 \ ,
\end{align}
where
\begin{align} \label{eq:lnz}
\ln Z_0 \equiv & - \frac{d}{2} \ln |B| - \frac{1}{2} \ln |P|_+ \nn \\
\ln Z_1 \equiv & \frac{g^2}{2} \Tr(P) + \sum_{a_i > 0} \ln D(a_i) + \kappa \ ,
\end{align}
and $\kappa$ is a function of $N_\rho$, $M$ and $g$, whose exact form is immaterial at the moment. \\

\noindent 1. {\it The connectivity}

The connectivity expectation value is obtained by
\begin{align}
\Cexp_a = \tau^2 \frac{\partial}{\der \tau} \left(\ln Z_0 + \ln Z_1 \right) \equiv \Cexp_0 + \Cexp_1 \ .
\end{align}
Starting with $\Cexp_0$, both terms of $\ln Z_0$ are proportional to a power of $\tau$:
\begin{align*}
|B| \sim \tau^{-N_\rho} \hs , \hs |P|_+ \sim \tau^{d (M - 1)} \ .
\end{align*}
and, using the fact that if $f (x) \sim x^\alpha$ then
\begin{align} \label{eq:derivative}
\frac{\der \ln f}{\der x} = \frac{\alpha}{x} \ ,
\end{align}
we obtain
\begin{align} \label{eq:cexp0}
\Cexp_0 = \frac{d \tau}{2} (N_\rho - M + 1) \ .
\end{align}
To calculate $\Cexp_1$, we note that $\ln Z_1$ depends on $\tau$ through $\Tr(P) \sim \tau$ and $a_i \sim \sqrt{p_i} \sim \sqrt{\tau}$. One can also obtain from eq. (\ref{eq:dawson}):
\begin{align} \label{eq:D_der}
\frac{\der \ln D(a_i)}{\der a_i} = \frac{1}{D(a_i)} - 2 a_i \ ,
\end{align}
and thence
\begin{align} \label{eq:cexp1}
\Cexp_1 &= \tau^2 \left\{ \frac{g^2}{2 \tau} \Tr(P) + \sum_{a_i > 0} \left[ \frac{1}{D(a_i)} - 2 a_i \right] \frac{a_i}{2 \tau} \right\} \nn \\
&= \frac{\tau}{2} \sum_{a_i > 0} \frac{a_i}{D(a_i)} \ ,
\end{align}
where the first and last terms in the first line cancel out. Summing (\ref{eq:cexp0}) and (\ref{eq:cexp1}) yields $\Cexp_a$ in eq. (\ref{eq:cexp}). \\

\noindent 2. {\it The volume}

The mean volume is obtained from eq. (\ref{eq:vexp_orig}) via
\begin{align}
\Vexp_a = - 2 W_{jk} \frac{\der \ln Z_a}{\der B_{jk}} + 2 U_{jk} \frac{\der \ln Z_a}{\der P_{jk}} \ .
\end{align}
Using again (\ref{eq:lnz}), we express a corresponding separation: $\Vexp_a = \Vexp_0 + \Vexp_1$. Starting with $\Vexp_0$ and noting that the dependencies on $B_{jk}$ and $P_{jk}$ originate in $|B|$ and $|P|_+$, respectively, we have
\begin{align*}
\Vexp_0 = - 2 W_{jk} \frac{\der \ln Z_0}{\der |B|} \frac{\der |B|}{\der B_{jk}} + 2 U_{jk} \frac{\der \ln Z_0}{\der |P|_+} \frac{\der |P|_+}{\der P_{jk}} \ .
\end{align*}
Using eq. (\ref{eq:derivative}) again we have:
\begin{align} \label{eq:vexp_mid}
\Vexp_0 = \frac{d W_{jk}}{|B|} \frac{\der |B|}{\der B_{jk}} - \frac{U_{jk}}{|P|_+} \frac{\der |P|_+}{\der P_{jk}} \ .
\end{align}
To evaluate these expressions, we express the determinants, e.g. $|B|$, as $|B| = \sum C_{jk} B_{jk}$, which leads to
\begin{align} \label{eq:det_der}
\frac{\der |B|}{\der B_{jk}} = C_{jk} = |B| (B^{-1})_{kj} \ ,
\end{align}
with $B^{-1}$ the inverse matrix of $B$. Similarly, for the pseudo-determinant $|P|_+$ we have
\begin{align} \label{eq:pseudo_det_der}
\frac{\der |P|_+}{\der P_{jk}} = |P|_+ \left(\Pi \right)_{kj} \ ,
\end{align}
with $\Pi$ the pseudo-inverse of $P$. Substituting (\ref{eq:det_der}) and (\ref{eq:pseudo_det_der}) into (\ref{eq:vexp_mid}) we obtain
\begin{align} \label{eq:vexp0}
\Vexp_0 = d \Tr(W B^{-1}) - \Tr(U \Pi) \ .
\end{align}

Turning to $\Vexp_1$, $\ln Z_1$ depends on $P_{jk}$ through $\Tr(P)$ and the eigenvalues $p_i$. The former gives $\der \Tr(P) / \der P_{jk} = \delta_{jk}$. For the latter, note that, if $O$ is the orthogonal diagonalisation matrix of $P$, $\left(O P O^T\right)_{ij} = \Lambda_{ij} = p_i \delta_{ij}$, then \cite{Ne76}
\begin{align} \label{eq:eigval_by_entry}
\frac{\der p_i}{\der P_{jk}} = O_{ij} O_{ik} \ ,
\end{align}
which also holds for non-zero eigenvalues of a singular matrix. Together with (\ref{eq:D_der}), these give
\begin{align} \label{eq:vexp1}
\Vexp_1 &= g^2 U_{jk} \delta_{jk} + 2 U_{jk} \sum_{a_i > 0} \left(\frac{1}{D(a_i)} - 2 a_i \right) \frac{a_i O_{ij} O_{ik}}{2 p_i} \nn \\
&= g^2 \Tr(U) + \sum_{a_i > 0} \left( \frac{1}{2 a_i D(a_i)} - 1 \right) g^2 U_{ii}' \nn \\
&= \sum_{a_i > 0} \frac{g^2 U_{ii}'}{2 a_i D(a_i)} \ ,
\end{align}
where we abbreviated $U_{jk} O_{ij} O_{ik} = (O U O^T)_{ii} \equiv U_{ii}'$. The first and last terms on the second line cancel out because $Tr(U)=Tr(U')$. Summing (\ref{eq:vexp0}) and (\ref{eq:vexp1}) gives $\Vexp_a$ in eq. (\ref{eq:vexp}). \\

\noindent 3. {\it The mean squared-norm of the force vector}

Noting that the expression
\begin{align}
\gexp_a = \frac{1}{Z_a} \int (g \cdot g) e^{- \frac{1}{2} \rhotv B \rhotv + \frac{1}{2} \gv P \gv} (d \rhotv)(d \gv)
\end{align}
has the same form as the volume expectation value in eq. (\ref{eq:vexp_orig}), with $W = \mathbb{0}$ and $U = \mathbb{1}$, we obtain $\gexp_a$ by substituting these values in eq. (\ref{eq:vexp}). This gives straightforwardly the result, eq. (\ref{eq:gexp}). \\

\noindent 4. {\it The entropy}

In analogy to thermal statistical mechanics, the affine entropy is given by
\begin{align} \label{eq:entropy_orig}
S_a = \frac{\Cexp_a}{\tau} + \ln Z_a \ .
\end{align}
Its calculation requires the explicit form of $\kappa$ in eq. (\ref{eq:lnz}),
\begin{align} \label{eq:const}
\kappa = \frac{d}{2} \left[ N_\rho \ln(2 \pi) + (M - 1) \ln(8) + 2 \ln (2 g) \right] \ .
\end{align}
To obtain $S_a$ in eq. (\ref{eq:entropy}), we substitute eq. (\ref{eq:lnz}), (\ref{eq:cexp}) and (\ref{eq:const}) into eq. (\ref{eq:entropy_orig}). \\

\noindent 5. {\it The boundary normal stress}

To obtain an explicit expression for $\sexp_a$ we must calculate it directly from the partition function. We shall use the following calculation:
\begin{align} \label{eq:single_int}
\frac{\int_0^g g' e^{\frac{1}{2} p g'^2} d g'}{\int_0^g e^{\frac{1}{2} p g'^2} d g'} = \frac{\left( e^{a^2} - 1 \right)}{\sqrt{2p} \ e^{a^2} D(a)} = \frac{g(1 - e^{-a^2})}{2 a D(a)} \ ,
\end{align}
where $p \geq 0$ and $a \equiv g \sqrt{p / 2}$. In the limit $p \to 0$ (\ref{eq:single_int}) reduces to $g/2$. Using this result, we define
\begin{align} \label{eq:curly_d}
\G_j \equiv \frac{1 - e^{-a_j^2}}{2 a_j D(a_j)} \cdot g
\end{align}
and proceed to calculate the expectation value of the normal component of a boundary force,
\begin{align}
\gixexp_a = \frac{1}{Z_a} \int e^{- \frac{1}{2} \rhotv B \rhotv} (d \rhotv) \int g^i_x e^{+ \frac{1}{2} \gv P \gv} (d \gv) \ .
\end{align}
The first set of integrals cancels out with the corresponding integrals of $Z_a$. To solve the second set of integrals, we diagonalise $P$ and use (\ref{eq:single_int}) and (\ref{eq:curly_d}) to obtain
\begin{align}
\gixexp_a = \sum_j O^T_{ij} \frac{\int g'_j e^{\frac{1}{2} p_j {g'_j}^2} d g'_j}{\int e^{\frac{1}{2} p_j {g'_j}^2} d g'_j} = \sum_j O^T_{ij} \G_j \ ,
\end{align}
with $O$ the diagonalisation matrix. Squaring,
\begin{align}
\gxexp_a^2 = \frac{1}{M} \sum_i \gixexp_a^2 = \frac{1}{M} \sum_{i,j,k} O^T_{ij} O^T_{ik} \G_j \G_k \ ,
\end{align}
and using the orthogonality, $\sum_i O^T_{ij} O^T_{ik} = \delta_{jk}$, we get
\begin{align} \label{eq:g_x_exp}
\gxexp_a^2 = \frac{1}{M} \sum_{j} \G_j^2 = \frac{|\G|^2}{M} \ ,
\end{align}
where $|\G|$ is the norm of a $d M$-long vector, whose components are $\G_j$. Combining (\ref{eq:g_x_exp}) with $\sexp_a = M \gxexp_a / 4 \sqrt{N} a$, as discussed in the main text, gives relation (\ref{eq:sexp}).

\section{Approximations of the Dawson functions} \label{appendix:approx}

The partition function contains $d (M - 1)$ Dawson integrals, which give rise to cumbersome sums of Dawson functions, $D(a_i)$, in the entropy (\ref{eq:entropy}) and the expectation values (\ref{eq:cexp}), (\ref{eq:gexp}) and (\ref{eq:vexp}). In this appendix, we approximate these sums in the limits of large and small $a_i= g \sqrt{p_i / 2} \sim g \sqrt{\tau} \gamma_j$, with $\gamma_{j=1,2}$ the two eigenvalues of the inverse angoricity $\gamma$. It is convenient to define the dimensionless parameter
\begin{align} \label{eq:}
\chi \equiv g \sqrt{\tau |\gamma|} \ ,
\end{align}
whose limits correspond to the limits of the parameters $a_i$.

\begin{figure}[h]
\includegraphics[width=0.45\textwidth]{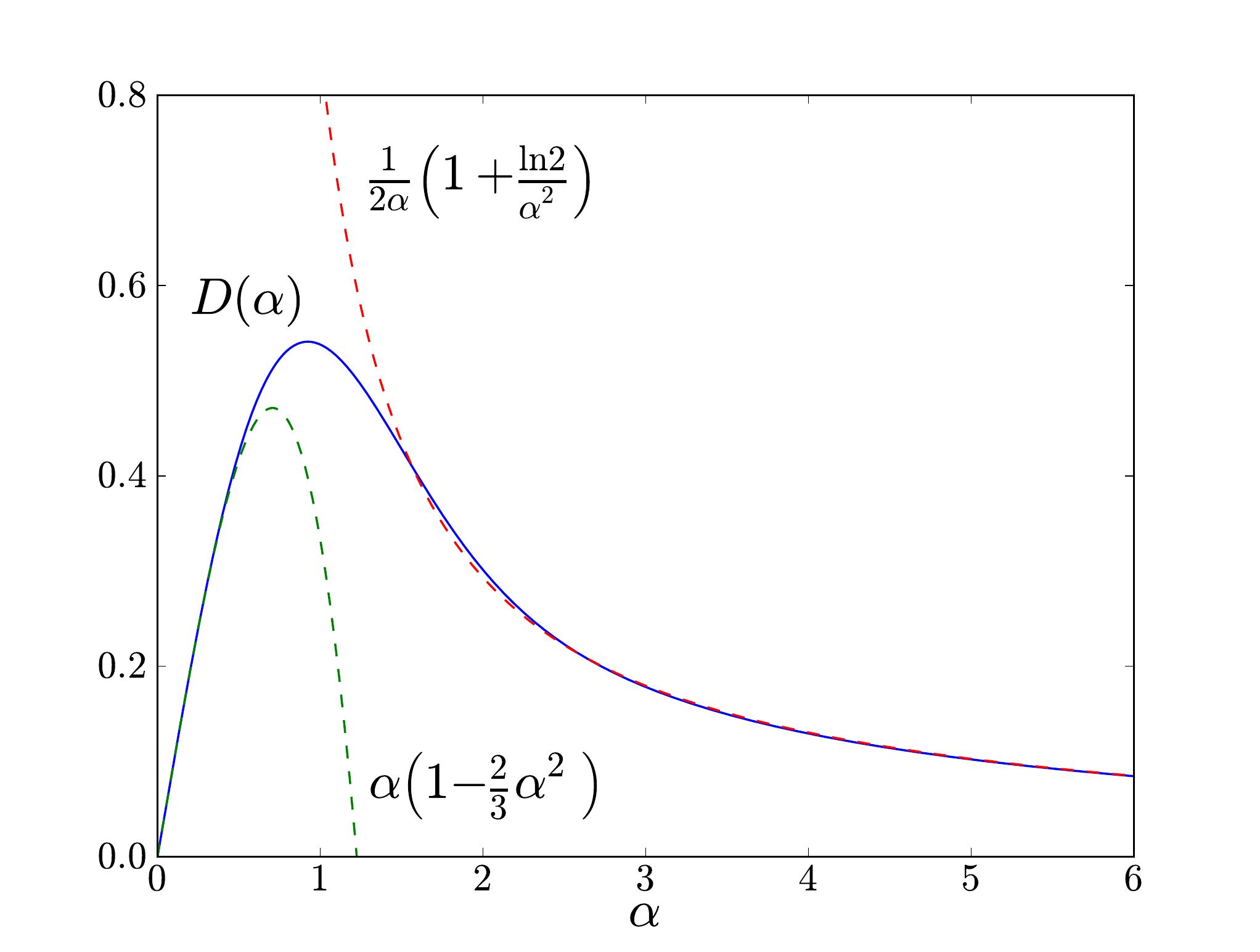}
\caption{The Dawson function, in solid blue line, alongside its small- and large-parameter two-term approximations, in green and red dashed lines, respectively. The accuracy of the respective approximations is better than 3\% outside the range $0.5 < \alpha < 1.5$.}
\label{fig:dawson_approx}
\end{figure}

The second order approximations of $D(\alpha)$, for small and large values of $\alpha$, are, respectively,
\begin{alignat}{3}
D(\alpha \ll 1) &= \hspace{0.15cm} \alpha &&\left(1 - \frac{2}{3}\alpha^2\right) \ , \label{SmallApprox} \\
D(\alpha \gg 1) &= \frac{1}{2\alpha} &&\left(1 + \frac{\ln 2}{\alpha^2}\right) \ . \label{LargeApprox}
\end{alignat}
In figure \ref{fig:dawson_approx} we show these approximations and observe that they are very accurate, except in the range $0.5 < \alpha < 1.5$, with a relative error of less than 3\%. The arguments of the Dawson function, $a_i$, are proportional to the maximal magnitude of a boundary force, $g$, to $\sqrt{\tau}$, and to the inverse angoricity eigenvalues, $\gamma_{1,2}$. Thus, for a given values of $g$ and $\tau$, the values of $a_i$s are small at high angoricity and vice versa. Nevertheless, we also need to find a good approximation for the regime $a_i \approx 1$, which we do next.

\noindent 1. {\it Approximation of the mean volume}

Here, we wish to approximate the sum $\sum 1 / (a_i D(a_i))$ in eq. (\ref{eq:vexp}). From relations (\ref{SmallApprox}) and (\ref{LargeApprox}) we observe that the sum is dominated by the small $a_i$s. We also note from \cite{Caetal97, *Rietal12} that, for a typical 2D system and $g \gamma_{1,2} \sqrt{\tau} = 1$ we get $0.5 \lesssim a_i \lesssim 200$, with most $a_i$s smaller than 1. Using then the small-parameter approximation we have:
\begin{align}
\sum_{a_i > 0} \frac{g^2 U_{ii}'}{2 a_i D(a_i)} &\approx \sum_{a_i > 0} \frac{g^2 U_{ii}'}{2 a_i^2} \left(1 + \frac{2}{3} a_i^2 \right) \nn \\
&= \sum_{a_i > 0} \frac{U_{ii}'}{p_i} + \frac{g^2}{3} \sum_{a_i > 0} U_{ii}' \nn \\
&= \Tr(U \Pi) + \frac{g^2}{3} \Tr(U \Pi P) \ .
\end{align}
For the last step we used again the diagonalisation of $\Pi$ to establish that the first sum is equal to $\Tr(U' \Lambda^{-1}) = \Tr(U \Pi)$ and the second sum is equal to $\Tr(\Lambda^{-1} \Lambda U') = \Tr(U \Pi P)$. Plugging these into eq. (\ref{eq:vexp}) we obtain for the mean volume
\begin{align} \label{eq:vexp_approx}
\Vexp_a \approx d \Tr(W B^{-1}) + \frac{g^2}{3} \Tr(U \Pi P) \ .
\end{align}

\noindent 2. {\it Approximation of the mean squared-norm of force vector}

An approximation of expectation value for the boundary forces can be obtained either by following the same route as above or by substituting $W = \mathbb{0}$ and $U = \mathbb{1}$ in eq. (\ref{eq:vexp_approx}), which yields
\begin{align} \label{eq:gexp_approx}
\gexp_a \approx \frac{g^2}{3} \Tr(\Pi P) = \frac{g^2}{3} d (M - 1) \ .
\end{align}
This is again a second order approximation, to first order, $-\Tr(\Pi)$ of eq. (\ref{eq:gexp}) cancels out with the sum to give, not surprisingly, $\gexp_a = 0$. This expression is independent of the contactivity, as one would expect intuitively. Its independence of the angoricity is only a feature of the leading term -- the next term would be proportional to $\Tr(\Pi P P)$, scaling as $\tau Tr(\gamma^2)$. \\

\noindent 3. {\it Approximations of the mean boundary normal stress and equation of state}

For these approximations, we use eqs. (\ref{eq:sexp}), (\ref{eq:curly_d}) and (\ref{eq:g_x_exp}).
Approximating $\G_j$ for small $a_j$ we have
\begin{align} \label{eq:curly_d_small}
\G_j = \frac{g}{2 a_j} \cdot \frac{1 - e^{-a_j^2}}{D(a_j)} \approx \frac{g}{2 a_j} \cdot \frac{a_j^2}{a_j} = \frac{g}{2} \ .
\end{align}
Substituting this in eq. (\ref{eq:sexp}) we get
\begin{align} \label{eq:sexp_approx}
\sexp_a \approx \frac{\sqrt{M}}{4 \sqrt{N} a} \sqrt{M \left(\frac{g}{2} \right)^2} \approx \frac{g}{8 a} \ .
\end{align}
Using the approximations for $\Vexp_a$, eq. (\ref{eq:vexp_approx}), and $\sexp_a$, eq. (\ref{eq:sexp_approx}), we obtain the second order approximation for the equation of state in the limit of low $\chi$
\begin{align} \label{eq:EOS_approx}
\Vexp_a \sexp_a \approx \frac{g}{8 a} \left[ d \Tr(W B^{-1}) + \frac{g^2}{3} \Tr(U \Pi P) \right] \ .
\end{align}

\noindent 4. {\it Approximation of the entropy}

\begin{figure}[h]
\includegraphics[width=0.45\textwidth]{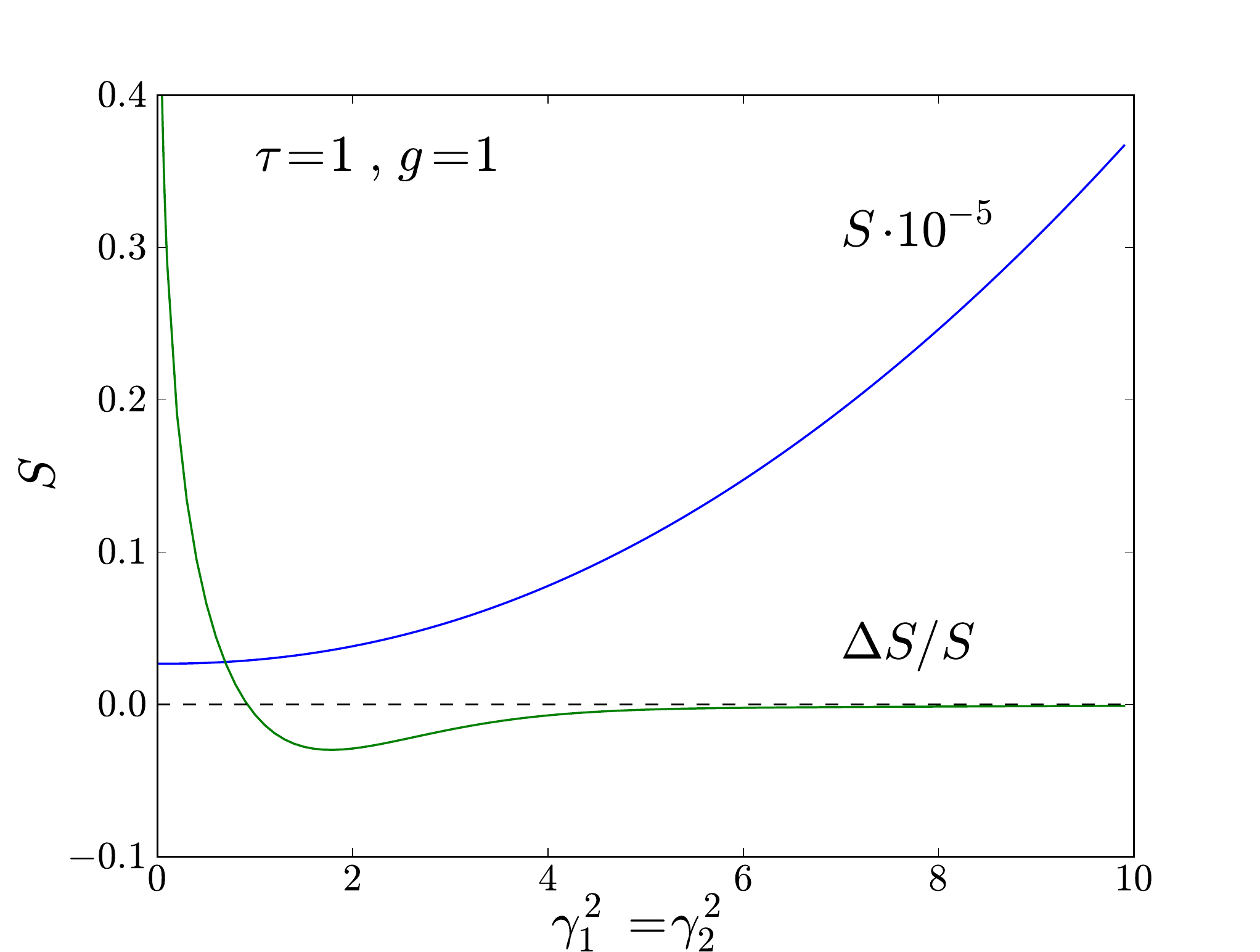}
\caption{{\it Blue, thin line:} the exact affine entropy, calculated according to eq. (\ref{eq:entropy}), for a 2D system of 1172 discs, as a function of the two squared eigenvalues (fixed to be equal) of $\gamma$. The contactivity, $\tau$, and the maximal boundary force, $g$, are set to 1. {\it Green, thick line:} the relative error of approximating the Dawson functions, namely the difference between calculating $S$ according to eq. (\ref{eq:entropy_approx}) and according to eq. (\ref{eq:entropy}), over the latter.}
\label{fig:entropy_approx}
\end{figure}

To this end, we need to approximate the expression $\Delta \equiv \sum [\ln D(a_i) + a_i / 2 D(a_i)]$ in eq. (\ref{eq:entropy}). Using relations (\ref{SmallApprox}) and (\ref{LargeApprox}), for small $a_i$'s, $\Delta \approx 0.5 + \ln a_i$ and for large $a_i$'s $\Delta \approx a_i^2 - \ln (4 a_i)$. It follows that large $a_i$'s dominate the sum. Moreover, using the definition of $a_i$, the term $\frac{g^2}{2} \Tr(P)$ in eq. (\ref{eq:entropy}) is also proportional to $\sum a_i^2$, making the large $a_i$ limit even more dominant. Using then (\ref{LargeApprox}), we have
\begin{align}
\Delta = &\sum_{a_i > 0} \left[\ln D(a_i) + \frac{a_i}{2 D(a_i)} \right] \nn \\
\approx &\sum_{a_i > 0} \left[a_i^2 - \ln a_i - \ln 4 \right] \nn \\
= &\frac{g^2}{2} \Tr(P) - \frac{1}{2} \ln |P|_+ - \frac{d}{2} (M - 1) \ln 8 g^2 \ .
\end{align}
Substituting this in eq. (\ref{eq:entropy}) we get:
\begin{align} \label{eq:entropy_approx}
S_a \approx &- \frac{d}{2} \ln |B| - \ln |P|_+ + g^2 \Tr(P) \\
&+ \frac{d}{2} \left[ N_\rho \ln (2 \pi e) + 2 \ln (2 g) - (M - 1) \ln (e g^2) \right] \ . \nn
\end{align}
The accuracy of this approximation depends on the different noise parameters through the dimensionless parameter $\chi$. Whenever $\chi \gtrsim 1$, the approximation (\ref{eq:entropy_approx}) is accurate to more than 3\%. This is demonstrated in figure \ref{fig:entropy_approx} for $\gamma_1 = \gamma_2$ and $\tau = g = 1$ - it shows the relative error, $(S_{\rm approx} - S) / S$, between the approximated entropy, eq. (\ref{eq:entropy_approx}), and the exact result, eq. (\ref{eq:entropy}), for different values of $\chi^2$. We find similar graphs when varying $\tau$, $g$ or the ratio of the eigenvalues of $\gamma$.

\section{Estimation of the boundary normal stress} \label{appendix:order_of_curly_d}

\begin{figure}[h]
\includegraphics[width=0.45\textwidth]{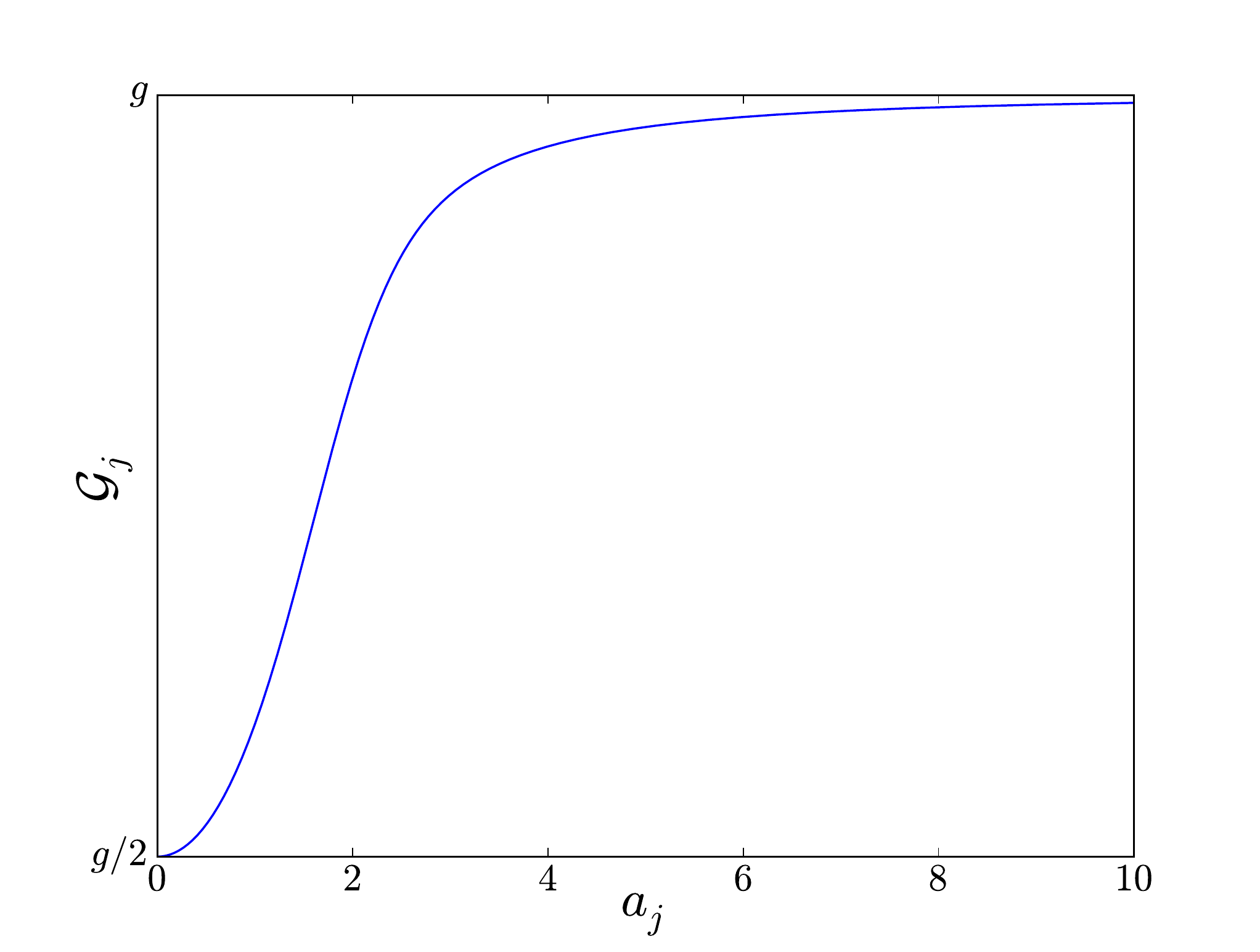}
\caption{A numerical check that indeed $\G_j(a_j)$ is bounded between $g/2$ and $g$ for all $j$.}
\label{fig:d_vs_a}
\end{figure}

Here, we first establish that $g/2 < \G_j < g$ for all $j$ and, thence, that $|\G| = O(\sqrt{M})$, as stated in section \ref{sec:calculations}. 
Using the definition of $\G(a_i)$ and the small $\chi$ approximation for $D(a_i)$, we have shown in eq. (\ref{eq:curly_d_small}) that $\G_j(a_j \ll 1) \approx g/2$. Similarly, using the large $\chi$ approximation, relation (\ref{LargeApprox}), we have
\begin{align}
\G_j(a_j \gg 1) \approx g\left(1 - \frac{\ln 2}{a_j^2}\right) < g \ .
\end{align}
This indicates that the value of $\G_j(a_j)$ is bounded between $g/2$ and $g$ for all $j$'s. We substantiate this by a calculation of the function $D_j(a_j)$, shown in figure \ref{fig:d_vs_a}. Thus, $g \sqrt{M} / 2 < |\G| < g \sqrt{M}$, and we obtain that, generally, $|\G| = O(M)$.

\end{appendices}

\newpage

\bibliography{FullEnsemblePaper}

\end{document}